\RequirePackage{fix-cm}

\documentclass[smallextended]{svjour3}       
\smartqed  
\usepackage{graphicx}
\usepackage{hyperref}
\usepackage{amsmath}
\usepackage{cite}
\usepackage{bm}
\usepackage{amssymb}
\usepackage{comment}

\begin{document}

\title{Renormalized stress-energy tensor on global anti-de Sitter space-time with Robin boundary conditions}

\author{Thomas Morley        \and Sivakumar Namasivayam \\ \and 
        Elizabeth Winstanley 
}

\institute{Consortium for Fundamental Physics, School of Mathematics and Statistics, The University of Sheffield,  Hicks Building, Hounsfield Road, Sheffield. S3 7RH United Kingdom
       \\       \email{Tom.Morley1994@gmail.com, SNamasivayam1@sheffield.ac.uk, E.Winstanley@sheffield.ac.uk}           
}

\date{\today }

\maketitle

\abstract{
We study the renormalized stress-energy tensor (RSET) for a massless, conformally coupled scalar field on global anti-de Sitter space-time in four dimensions.
Robin (mixed) boundary conditions are applied to the scalar field.
We compute both the vacuum and thermal expectation values of the RSET. 
The vacuum RSET is a multiple of the space-time metric when either Dirichlet or Neumann boundary conditions are applied.
Imposing Robin boundary conditions breaks the maximal symmetry of the vacuum state and results in an RSET whose components with mixed indices have their maximum (or maximum magnitude) at the space-time origin. 
The value of this maximum depends on the boundary conditions. 
We find similar behaviour for thermal states. 
As the temperature decreases, thermal expectation values of the RSET approach those for vacuum states and their values depend strongly on the boundary conditions.
As the temperature increases, the values of the RSET components tend to profiles which are the same for all boundary conditions.
We also find, for both vacuum and thermal states, that the RSET on the space-time boundary is independent of the boundary conditions and determined entirely by the trace anomaly.}


\maketitle

\section{Introduction}
\label{sec:intro}
In the absence of a full theory of quantum gravity, quantum field theory in curved space-time (QFTCS) provides us with an effective theory in which we study quantum fields propagating on a background classical curved space-time. 
In QFTCS,  the renormalized expectation value of the stress-energy tensor operator (RSET)  $\langle \hat{T}_{\mu\nu}\rangle $ plays a pivotal role. The expectation value of this operator is used as the source term in  the semiclassical version of Einstein's field equations~\eqref{eq:EFE} (here and through this paper we use units in which $\hbar=c=G=1$): 
\begin{equation}
    R_{\mu\nu}-\frac{1}{2}R \, g_{\mu\nu}+ g_{\mu\nu} \Lambda = 8\pi  \langle \hat{T}_{\mu\nu}\rangle ,
    \label{eq:EFE}
\end{equation}
and therefore governs the backreaction effect of the quantum field on the space-time geometry.

In this paper we consider the RSET for a quantum scalar field on global anti-de Sitter (adS) space-time.
Although this is a maximally symmetric space-time, quantum fields on this background have rich properties, not least because of the need to impose boundary conditions on the field due to the fact that adS is not a globally hyperbolic space-time.
The study of quantum fields on adS was initiated many years ago \cite{Avis:1977yn}, where a massless, conformally coupled scalar field was studied, subject to either ``transparent'' or ``reflective'' boundary conditions. 
The latter correspond to either Dirichlet (the field vanishes on the boundary) or Neumann (the normal derivative of the field vanishes on the boundary) boundary conditions. 
The vacuum state retains the maximal symmetry of the underlying geometry when either Dirichlet or Neumann boundary conditions are applied and the vacuum expectation value of the RSET is a constant multiple of the space-time metric \cite{Allen,Allen:1986,Kent:2014nya}. 
The introduction of a nonzero temperature breaks this symmetry but, nonetheless, the thermal expectation value of the RSET for a massless, conformally coupled scalar field can be found using an elegant method based on the time-periodicity properties of the thermal Green's function \cite{Allen:1986}.

The simplest boundary conditions, as studied in \cite{Avis:1977yn,Allen,Allen:1986,Kent:2014nya,Ambrus:2018olh}, are by no means the only possibilities \cite{Avis:1977yn,Barroso:2019cwp,Dappiaggi:2017wvj,Dappiaggi:2018xvw,Dappiaggi:2018pju,Ishibashi:2004wx,Benini:2017dfw,Ishibashi:2003jd,Wald:1980jn,Dappiaggi:2021wtr,Gannot:2018jkg,Campos:2022byi}.
The wide range of valid boundary conditions gives rise to an extensive set of possible quantum states that can be studied. 
Amongst the various possible boundary conditions, in this paper we focus on Robin (mixed) boundary conditions (see, for example, \cite{Dappiaggi:2017wvj,Dappiaggi:2018pju,Dappiaggi:2021wtr,Dappiaggi:2022dwo,Campos:2022byi} for more general boundary conditions that can be applied).
For a massless, conformally coupled scalar field, Robin boundary conditions correspond to the vanishing of a linear combination of the field and its normal derivative on the boundary.
Such boundary conditions break the maximal symmetry of the vacuum state \cite{Barroso:2019cwp,Dappiaggi:2016fwc,Dappiaggi:2018xvw,Dappiaggi:2021wtr,Pitelli:2019svx}.

The renormalized vacuum polarization (VP, the square of the scalar field) was computed in \cite{Morley:2021} for a massless, conformally coupled scalar field on four-dimensional adS with Robin boundary conditions applied to all field modes.
For both vacuum expectation values (v.e.v.s) and thermal expectation values (t.e.v.s) it was found that, on the space-time boundary, the VP has the same value for all boundary conditions except for Dirichlet, where the value was different. 
The same conclusion was reached recently \cite{Namasivayam:2022} on three-dimensional adS for a scalar field with nonzero mass and values of the coupling to the space-time curvature for which Robin boundary conditions can be applied.
As a result, while Dirichlet boundary conditions are the most widely considered in the literature due to their simplicity, it is the Neumann boundary conditions which give the generic behaviour of the VP on the space-time boundary.
In contrast, if Robin boundary conditions are applied only to a subset of the scalar field modes corresponding to $s$-wave perturbations, then the VP for a massless, conformally coupled scalar field on four-dimensional adS
always takes the Dirichlet value on the boundary \cite{Barroso:2019cwp}.

In this paper we explore whether the result in \cite{Morley:2021} extends to the v.e.v.s and t.e.v.s of the RSET for a massless, conformally coupled scalar field on four-dimensional adS.
In \cite{Barroso:2019cwp}, applying Robin boundary conditions just to the $s$-wave modes, it is found that the RSET on the space-time boundary again takes the same value as for Dirichlet boundary conditions.
Here we follow \cite{Morley:2021} and apply Robin boundary conditions to {\em {all}} field modes.
As in \cite{Morley:2021}, we employ Euclidean methods to find the v.e.v.s and t.e.v.s of the RSET, paying particular attention to how these depend on the parameter describing the Robin boundary conditions.

Our paper is structured as follows. 
In Section~\ref{sec:Euclid} we outline the construction of the vacuum and thermal Green's functions for a massless, conformally coupled, scalar field on four-dimensional adS. 
This is followed, in Section~\ref{sec:EXPSET}, with the calculation of the expectation values for the RSET in both vacuum and thermal states, including a brief discussion of the numerical methods employed.
The results for the v.e.v.s and t.e.v.s for the RSET are given in Sections~\ref{sec:vev with Robin} and~\ref{sec:tev with Robin} respectively. 
The behaviour of these quantities approaching the space-time boundary is explored further in Section~\ref{sec:boundary}.
Finally we present our conclusions in Section~\ref{sec:discussion}.

\section{Euclidean Green's functions}
\label{sec:Euclid}
AdS space-time is a maximally symmetric solution of Einstein's field equations of general relativity, with constant negative curvature. In global coordinates $(t,\rho, \theta ,\phi )$ the metric is 
\begin{equation}
ds^2=L^2 \sec^2\rho \,[-dt^2 + d \rho^2 + \sin^2\rho \, ( d\theta^2 + \sin^2\theta \, d\phi^2)],
\label{eq:adsmetric}
\end{equation}
where $ \,0\le \rho <\pi/2,\, 0\le \theta <\pi,\,\text{and}\, 0 \le\phi < 2\pi$. In four dimensions, the cosmological constant ($\Lambda <0$) is related to  the adS radius of curvature, $L$,  via $\Lambda = -3/L^2$. In adS the time coordinate is periodic with  $t \in (-\pi,\pi]$  and the end points identified. This results in somewhat unphysical closed time-like curves. This is circumvented by considering the covering space  (CadS) where the time coordinate is `unwrapped' to give $ -\infty < t < \infty$.

We work in Euclidean space where the Green's function is a unique, well-defined distribution.
The Euclidean  metric is obtained from the adS metric 
\eqref{eq:adsmetric} by performing a Wick rotation, $t \to i \tau$, leading to
\begin{equation}
ds^2 = L^2\sec^2\rho \, \,[d\tau^2 + d\rho^2 +\sin^2\rho \, ( d\theta^2 +\sin^2\theta \, d\phi^2)].
\label{eq:Euclid_metric}
\end{equation}
We consider the Euclidean Green's functions for a massless, conformally coupled scalar field in the vacuum state and in a  thermal state at inverse temperature $\beta $. 
The vacuum Green's function $G_{\zeta ,0}^{\rm {E}}(x,x')$ takes the form \cite{Morley:2021}
\begin{equation}
G_{\zeta ,0}^{\rm {E}}(x,x')  =\frac{1}{8\pi^2L^2}\cos \rho \cos \rho'
   \int _{\omega=-\infty}^\infty d\omega\, e^{i \omega \Delta \tau} \sum_{\ell=0}^\infty (2\ell +1) P_\ell(\cos\gamma) g_{\omega \ell} (\rho,\rho'),
\end{equation}
where $\omega $ is the frequency, $\Delta \tau = \tau - \tau'$, the radial Green's function is denoted by $g_{\omega \ell}(\rho , \rho')$ and $P_\ell(x)$ is a  Legendre polynomial.
The  angular separation of the space-time points, $\gamma$, is given by 
\begin{equation}
    \cos \gamma= \cos \theta \cos \theta' + \sin \theta \sin\theta' \cos \Delta \phi,
    \label{eq:gamma}
\end{equation}
where $\Delta \phi = \phi - \phi '$.
For a thermal state at inverse temperature $\beta $, the frequency $\omega $ takes the quantized values $\omega = n\kappa $ where $\kappa$ is related to the inverse temperature by
\begin{equation}
    \kappa= \frac{2 \pi}{\beta } .
    \label{eq:kappa}
\end{equation}
With this notation, the thermal Euclidean Green's function $G_{\zeta ,\beta }^{\rm {E}}(x,x')$ is then \cite{Morley:2021}  
\begin{equation}
    G_{\zeta,\beta}^{\rm {E}}(x,x')  = \frac{\kappa}{8\pi^2L^2} \cos \rho\, \cos \rho' \sum^\infty _{n=-\infty} e^{i n \kappa \Delta \tau} \sum_{\ell=0}^\infty (2\ell +1) P_\ell(\cos\gamma) g_{\omega \ell} (\rho,\rho') .
\end{equation}

The radial Green's function $g_{\omega \ell}(\rho ,\rho')$  satisfies the inhomogenous equation
\begin{equation}
    \left \{ \frac{d}{d\rho}\left ( \sin^2 \rho \frac{d}{d\rho}\right)
    -\omega^2\sin^2\rho - \ell(\ell+1)\right \} g_{\omega \ell}(\rho, \rho')= \delta(\rho-\rho') ,
    \label{eq:radeqn}
\end{equation}
and takes the form 
\begin{equation}
    g_{\omega \ell} (\rho, \rho') = \frac{p_{\omega \ell}(\rho_<) q_{\omega \ell}(\rho_>)}{{N}_{\omega \ell}},
\end{equation}
where $\rho_< = \text{min}\{\rho,\rho'\}$ and $\rho_> = \text{max}\{\rho, \rho'\}$,  with ${N}_{\omega \ell}$ a normalization constant.  
Here $p_{\omega \ell }$ and $q_{\omega \ell }$ are solutions of the homogeneous version of (\ref{eq:radeqn}) and can be written in terms of conical (Mehler) functions. 
The function $p_{\omega \ell }(\rho )$ is regular at the origin $\rho =0$ and takes the form
\begin{equation}
    p_{\omega \ell}(\rho)= \frac{1}{\sqrt{\sin\rho}} P_{i\omega -1/2}^{-\ell-1/2}(\cos \rho), 
\end{equation}
where $P_\mu^\nu(z)$ are associated Legendre functions.
At $\rho = \pi /2$, the function $q_{\omega \ell }(\rho )$ satisfies Robin boundary conditions:
\begin{equation}
    q_{\omega \ell}(\rho)   \cos \zeta + \frac{d q_{\omega \ell}(\rho)}{d \rho} \sin \zeta=0,
    \label{eq:Robin}
\end{equation}
 where $\zeta \in [0,\pi)$ is the Robin parameter. 
 The value $\zeta =0$ corresponds to Dirichlet boundary conditions, while $\zeta= \pi/2$ gives Neumann boundary conditions.
 Imposing (\ref{eq:Robin}) on the general solution of the homogeneous version of (\ref{eq:radeqn}) gives
 \begin{equation}
    q_{\omega \ell}= \frac{1}{\sqrt{\sin \rho}}\left[ C_{\omega \ell}^\zeta P_{i\omega -1/2}^{-\ell-1/2}(\cos \rho) + P_{i\omega -1/2}^{-\ell-1/2} (-\cos\rho)\right],
\end{equation}
where the constant $C_{\omega \ell}^\zeta$ is given by 
\begin{equation}
    C_{\omega \ell}^\zeta=\frac{2|\Gamma(\frac{i\omega + \ell+2)}{2})|^2 \tan\zeta-|\Gamma(\frac{i\omega+\ell+1}{2})|^2}{2|\Gamma(\frac{i\omega +\ell+2}{2})|^2 \tan\zeta +|\Gamma (\frac{i\omega +\ell+1}{2})|^2}.
\end{equation}
We have $C_{\omega \ell }^{0}=-1$ for Dirichlet boundary conditions and $C_{\omega \ell }^{\pi /2}=1$ for Neumann boundary conditions.
The normalization constant $N_{\omega \ell}$ is then given by \cite{Morley:2021}
\begin{equation}
    N_{\omega \ell} = \frac{2}{|\Gamma(\ell + 1+ i\omega)|^2}.
\end{equation}

Following \cite{Morley:2021}, we now write the vacuum and thermal Euclidean Green's function with Robin boundary conditions as  follows: 
\begin{align}
G_{\zeta,0}^{\rm {E}} (x,x') &=G_{{\rm {D}},0}^{{\rm {E}}}(x,x')\cos^{2}\zeta +G_{{\rm {N}},0}^{\rm {E}}(x,x')\sin^{2}\zeta +G_{{\rm {R}},0}^{\rm {E}}(x,x')\sin 2\zeta , \label{eq:vac_therm_Green1} \\
G_{\zeta,\beta}^{\rm {E}}(x,x') &=G_{{\rm {D}},\beta}^{\rm {E}}(x,x')\cos^{2}\zeta +G_{{\rm {N}},\beta}^{\rm {E}}(x,x')\sin^{2}\zeta +G_{{\rm {R}},\beta}^{{\rm {E}}}(x,x')\sin 2\zeta,
\label{eq:vac_therm_Green}
\end{align}
where $G_{{\rm {D}},0}^{{\rm {E}}}(x,x')$ and $G_{{\rm {D}},\beta}^{\rm {E}}(x,x')$ are the vacuum and thermal Euclidean Green's 
functions with Dirichlet boundary conditions, given by 
    \begin{align} 
        G_{{\rm {D}},0}^{\rm {E}}(x,x') &= \frac{1}{16\pi^2L^2}\frac{\cos\rho\cos \rho'}{\sqrt{\sin\rho \sin\rho'}}\int_{\omega=-\infty}^\infty d\omega  \, e^{i \omega \Delta \tau} \sum _{\ell=0}^\infty (2\ell+1)
        P_\ell(\cos \gamma) |\Gamma(\ell+1+i\omega)|^2  \nonumber \\
        &  \qquad \times P_{i\omega -1/2}^{-\ell-1/2}(\cos\rho_<)\left[ 
        P_{i\omega-1/2}^{-\ell-1/2}(-\cos\rho_>)-P_{i\omega-1/2}^{-\ell-1/2}(\cos\rho_>)\right],
        \\
      G_{{\rm {D}},\beta}^{\rm {E}}(x,x') & = \frac{\kappa}{16\pi^2L^2}\frac{\cos\rho\cos \rho'}{\sqrt{\sin\rho \sin\rho'}}\sum_{n=-\infty}^\infty e^{in\kappa \Delta \tau} \sum _{\ell=0}^\infty (2\ell+1)
        P_\ell(\cos \gamma) |\Gamma(\ell+1+i n \kappa)|^2  \nonumber \\
        & \qquad \times P_{i n \kappa -1/2}^{-\ell-1/2}(\cos\rho_<)\left[ 
        P_{i n \kappa-1/2}^{-\ell-1/2}(-\cos\rho_>)-P_{i n\kappa-1/2}^{-\ell-1/2}(\cos\rho_>)\right] ,   
\end{align}
$G_{{\rm {N}},0}^{\rm {E}}(x,x')$ and $G_{{\rm {N}},\beta}^{\rm {E}}(x,x')$ are the vacuum and thermal Euclidean Green's  functions, with Neumann boundary conditions,  given by 
    \begin{align}
        G_{{\rm {N}},0}^{\rm {E}}(x,x') &= \frac{1}{16\pi^2L^2}\frac{\cos\rho\cos \rho'}{\sqrt{\sin\rho \sin\rho'}}\int_{\omega=-\infty}^\infty d\omega \, e^{i \omega \Delta \tau} \sum _{\ell=0}^\infty (2\ell+1)
        P_\ell(\cos \gamma) |\Gamma(\ell+1+i\omega)|^2  \nonumber \\
        &   \qquad \times P_{i\omega -1/2}^{-\ell-1/2}(\cos\rho_<)\left[ 
        P_{i\omega-1/2}^{-\ell-1/2}(-\cos\rho_>)+P_{i\omega-1/2}^{-\ell-1/2}(\cos\rho_>)\right],\\
      G_{{\rm {N}},\beta}^{\rm {E}}(x,x') &= \frac{\kappa}{16\pi^2L^2}\frac{\cos\rho\cos \rho'}{\sqrt{\sin\rho \sin\rho'}}\sum_{n=-\infty}^\infty e^{in\kappa \Delta \tau} \sum _{\ell=0}^\infty (2\ell+1)
        P_\ell(\cos \gamma) |\Gamma(\ell+1+i n \kappa)|^2 \nonumber  \\
        &  \qquad \times P_{i n \kappa -1/2}^{-\ell-1/2}(\cos\rho_<)\left[ 
        P_{i n \kappa-1/2}^{-\ell-1/2}(-\cos\rho_>)+P_{i n\kappa-1/2}^{-\ell-1/2}(\cos\rho_>)\right] ,  
    \end{align}
$G_{{\rm {R}},0}^{\rm {E}}(x,x')$ and $G_{{\rm {R}},\beta}^{{\rm {E}}}(x,x')$ are the vacuum and thermal regular contributions (not Green's functions), given  by 
\begin{align}
        G_{{\rm {R}},0}^{\rm {E}}(x,x') & =\frac{1}{16\pi^2L^2}\frac{\cos\rho\,\cos\rho'}{\sqrt{\sin\rho\,\sin\rho'}}\int_{\omega=-\infty}^{\infty} d \omega \, e^{i \omega \Delta \tau} \sum _{\ell=0}^\infty D_{\omega \ell }^{\zeta } P_\ell(\cos\gamma) 
        P^{-\ell-1/2}_{i\omega-1/2} (\cos \rho) \, P^{-\ell-1/2}_{i\omega-1/2} (\cos \rho'), 
        \label{eq:reg_vac}
\\
        G_{{\rm {R}},\beta}^{\rm {E}}(x,x') &=\frac{\kappa}{16\pi^2L^2}\frac{\cos\rho\,\cos\rho'}{\sqrt{\sin\rho\,\sin\rho'}}\sum_{n=-\infty}^{\infty} e^{i n\kappa \Delta \tau} \sum _{\ell=0}^\infty D_{\omega \ell }^{\zeta }P_\ell(\cos\gamma)
        P^{-\ell-1/2}_{in \kappa-1/2}( \cos \rho) \, P^{-\ell-1/2}_{i n \kappa-1/2} (\cos \rho') , 
        \label{eq:reg_therm}
\end{align}
where the constants $D_{\omega \ell }^{\zeta }$ are given by
\begin{equation}
D^{\zeta }_{\omega \ell } =(2\ell+1)|\Gamma(1+\ell+i\omega)|^2 \left[ \frac{2|\Gamma(\frac{i \omega  +\ell +2}{2})|^2 \cos \zeta-|\Gamma(\frac{i \omega +\ell +1}{2})|^2 \sin \zeta}{2|\Gamma(\frac{i \omega  +\ell +2}{2})|^2 \sin \zeta+|\Gamma(\frac{i \omega +\ell +1}{2})|^2 \cos \zeta} \right] ,
\label{eq:Dconstants}
\end{equation}
with $\omega = n\kappa $ in the thermal sum (\ref{eq:reg_therm}).
 It is clear from 
 (\ref{eq:reg_vac}, \ref{eq:reg_therm}), that $G_{{\rm {R}},0}^{\rm {E}}(x,x'), G_{{\rm {R}},\beta}^{\rm {E}}(x,x')$ will diverge if there are values of the Robin parameter $\zeta$ satisfying 
\begin{equation}
    2\tan \zeta =-\frac{|\Gamma(\frac{i\omega +\ell+1}{2})|^2}{|\Gamma(\frac{i\omega +\ell+2}{2})|^2}. 
    \label{eq:unstable}
\end{equation}
If  the Robin parameter $\zeta $ lies in the interval $\zeta_{\text{crit}}<\zeta < \pi $, where $\zeta_{\text{crit}}\approx 0.68 \pi $ for a massless, conformally coupled scalar field on four-dimenional adS, then there exist real values of  $\omega $ satisfying (\ref{eq:unstable}) \cite{Morley:2021}. 
Such values of $\omega $ give rise to classical mode solutions of the scalar field equation on the adS space-time (\ref{eq:adsmetric}) which are exponentially growing in time and are therefore classically unstable \cite{Morley:2021}. 
To consider a quantum scalar field, we require the classical scalar field to be classically stable,  so for the remainder of this work we restrict our consideration of Robin boundary conditions to values of the Robin parameter $\zeta $ in the interval $0 \le \zeta <\zeta_{\text{crit}}$.

\section{Expectation values of the RSET}
\label{sec:EXPSET}
Having given expressions for  the vacuum and thermal Green's functions for the massless, conformally coupled, scalar field on four-dimensional  adS, we now 
determine the  v.e.v.s and t.e.v.s of the RSET.
For both vacuum and thermal states, the Euclidean Green's functions are singular in the coincidence limit $x'\rightarrow x$. 
Assuming that both these states satisfy the Hadamard condition, the singular part of the Green's function is given by the singular part of the Hadamard parametrix $G^{\rm {S}}(x,x')$, which is independent of the quantum state. 
The RSET expectation value for a particular quantum state  is found by subtracting $G^{{\rm {S}}}(x,x')$ from the Euclidean Green's function $G^{\rm {E}}(x,x')$ for that state, applying a second order differential operator, and then taking the coincidence limit, namely~\cite{Decanini:2005eg}:
\begin{equation}
    \langle {\hat {T}}_{\mu\nu}(x)\rangle= \lim_{x'\to x}\left\{ \mathcal{T}_{\mu\nu}(x,x')\left[ G^{\rm {E}}(x,x') - G^{{\rm {S}}}(x,x') \right] - g_{\mu \nu }v_{1}(x,x') \right\},
    \label{eq:SETexp_def}
\end{equation} 
where 
$\mathcal{T}_{\mu\nu}(x,x')$ is the second order differential operator~\cite{Decanini:2005eg}
\begin{equation}
\begin{split}
    \mathcal{T}_{\mu\nu} & = \frac{2}{3}g_{\nu}^{\,\,\nu'}\nabla_\mu \nabla_{\nu'} - \frac{1}{6}g_{\mu \nu} g^{\rho \sigma'} \nabla_\rho \nabla_{\sigma'}
    -\frac{1}{3} g^{\,\,\mu'}_\mu g^{\,\,\nu'}_\nu \nabla_{\mu'}\nabla_{\nu'} 
   +\frac{1}{3} g_{\mu\nu} \nabla_\rho \nabla^\rho + \frac{1}{6} \Big( R_{\mu \nu} -\frac{1}{2} g_{\mu\nu} R \Big),
    \label{eq:Tmunuop}
    \end{split}
\end{equation}
and $g_{\mu\nu'}$ represents the bivector of parallel transport  between the points $x$ and $x'$.
The final term in (\ref{eq:SETexp_def}), $v_{1}(x,x')$, is a state-independent biscalar, regular in the coincidence limit, which ensures that the RSET is conserved~\cite{Decanini:2005eg}. 
Using (\ref{eq:vac_therm_Green1}, \ref{eq:vac_therm_Green}, \ref{eq:SETexp_def}) the vacuum/thermal  expectation values of the stress energy tensor $\langle {\hat {T}}_{\mu\nu}\rangle^\zeta$ can be written as:
\begin{multline}
\langle {\hat {T}}_{\mu\nu}\rangle^\zeta=\lim\limits_{x'\rightarrow x}\{\mathcal{T}_{\mu\nu}(x,x')\left[ G_{\rm {D}}^{\rm {E}}(x,x') - G^{\rm {S}}(x,x') \right] - g_{\mu \nu }v_{1}(x,x')  \}\cos^{2}\zeta
\\ +\lim\limits_{x'\rightarrow x}\left\{\mathcal{T}_{\mu\nu}(x,x') \left[ G_{{\rm {N}}}^{\rm {E}}(x,x') - G^{\rm {S}}(x,x') \right] - g_{\mu \nu }v_{1}(x,x')  \right\}\sin^{2}\zeta\\+\lim\limits_{x'\rightarrow x}\{\mathcal{T}_{\mu\nu}(x,x')G_{{\rm {R}}}^{\rm {E}}(x,x')\}\sin 2\zeta. 
\label{eq:total_SET_raw}
\end{multline}
The quantity $G_{{\rm {R}}}^{\rm {E}}(x,x')$ in the final term in \eqref{eq:total_SET_raw} is regular in the coincidence limit. 
Since the singular terms $\mathcal{T}_{\mu\nu}(x,x')G^{\rm {S}}(x,x') $ and the final subtraction term $g_{\mu \nu }v_{1}(x,x')$ are both independent of the quantum state, we therefore obtain the following expression for the expectation values of the stress-energy tensor when Robin boundary conditions are applied, in terms of those when Dirichlet or Neumann boundary conditions are applied:
\begin{equation}
\langle\hat{T}_{\mu\nu}\rangle^{\zeta}_{\text{ren }}=\langle\hat{T}_{\mu\nu}\rangle^{\rm {D}}_{\text{ren}} \cos^{2}\zeta+\langle\hat{T}_{\mu\nu}\rangle^{\rm {N}}_{\text{ren}}\sin^{2}\zeta+\lim\limits_{x'\rightarrow x}\{\mathcal{T}_{\mu\nu}(x,x')G_{\rm {R}}^{\rm {E}}(x,x')\}\sin 2\zeta.
\label{eq:SETexp_ren}
\end{equation}
From henceforth, the `ren' subscript will be omitted and it can be assumed that all $\langle \hat{T}_{\mu\nu} \rangle $ terms are renormalised. The t.e.v.s of the RSET with Dirichlet and Neumann boundary conditions have been determined in \cite{Allen:1986} (see (3.13), in which there is a minor typographical error which is corrected below):
\begin{align}
\langle\hat{T}_{\mu\nu}\rangle_{\beta}^{{\rm {D/N}}}=&\frac{1}{8\pi^2L^4}\left\{\left[-\frac{1}{120}+\frac{4}{3}\cos^{4}\rho \  f_{3}\left(\frac{\beta}{L}\right)\right]g_{\mu\nu}+\left[\frac{16}{3}\cos^{4}\rho\, f_{3}\left(\frac{\beta}{L}\right)\right]\tau _{\mu}\tau _{\nu}\right\}\nonumber\\
&\pm \frac{\cot\rho}{8\pi^2L^4} \left\{\left[-\frac{1}{6}\csc^{2}\rho \ \cos 2\rho \ S_{0}\left(\frac{\beta}{L},\rho\right)+\frac{1}{3}\cot\rho \ C_{1}\left(\frac{\beta}{L},\rho\right)+\frac{2}{3}\cos^{2}\rho \ S_{2}\left(\frac{\beta}{L},\rho\right)\right]g_{\mu\nu}\right.\nonumber\\
&\left.+\left[\frac{1}{6}(3-\cot^{2}\rho)S_{0}\left(\frac{\beta}{L},\rho\right)+\cot\rho\left(1-\frac{2}{3}\cos^{2}\rho\right)C_{1}\left(\frac{\beta}{L},\rho\right)+2\cos^{2}\rho \  S_{2}\left(\frac{\beta}{L},\rho\right)\right]\tau _{\mu}\tau _{\nu}\right.\nonumber\\
&\left.+\left[\frac{1}{6}(3\csc^{2}\rho-4)S_{0}\left(\frac{\beta}{L},\rho\right)+\cot\rho\left(\frac{2}{3}\sin^{2}\rho-1\right)C_{1}\left(\frac{\beta}{L},\rho\right)-\frac{2}{3}\cos^{2}\rho \ S_{2}\left(\frac{\beta}{L},\rho\right)\right]\rho_{\mu}\rho_{\nu}\right\} ,
\label{eq:SETexpDN}
\end{align}
where $g_{\mu \nu }$ is the space-time metric (\ref{eq:Euclid_metric}), and $\tau _{\mu }$, $\rho _{\mu }$ are unit vectors in the $\tau $ and $\rho $ directions, respectively.
The Dirichlet boundary condition corresponds to the $+$ sign whilst the Neumann boundary condition has the $-$ sign. The functions $f_m$, $S_m$ and $C_m$ are given by \cite{Allen:1986}
\begin{align}
    f_m(x)&=\sum_{n=1}^\infty n^m(e^{nx}-1)^{-1},  \\
    S_m(x, \rho)&= \sum_{n=1}^\infty n^m(-1)^n (e^{nx}-1)^{-1} \sin(2n\rho),  \\
    C_m(x,\rho)&=\sum_{n=1}^\infty n^m(-1)^n (e^{nx}-1)^{-1} \cos(2n\rho).
     \end{align} 
 To obtain  the corresponding vacuum expectation value from \eqref{eq:SETexpDN}, we take the limit as $\beta \to \infty$, as will be discussed in Section~\ref{sec:vev with Robin}.
 
The expression $\lim\limits_{x'\rightarrow x}\{\mathcal{T}_{\mu\nu}(x,x')G_{{\rm {R}},\beta}^{{\rm {E}}}(x,x')\}$ in the final term in \eqref{eq:SETexp_ren}, has been evaluated with {\small MATHEMATICA}. Using the recurrence relations for the conical functions \cite[\S 14.10.1]{NIST:DLMF}, the nonzero components of this contribution to the v.e.v.s  of the RSET are given by 
  {\allowdisplaybreaks \begin{align}
  \langle {\hat {T}}_\tau^\tau \rangle _{\text{R},0}^\zeta & =\frac{\cos\rho\cot^3\rho}{192L^4\pi^2} \sum_{\ell=0}^\infty\, \int_{\omega=-\infty}^{\infty} \, D_{\omega \ell}^\zeta \Bigg\{ -2\,\chi_{\omega \ell}^2 \,[P_{i\omega-1/2}^{-3/2-\ell} (\cos\rho)]^2 \,\sin^2\rho
  \nonumber \\ & \qquad
  - 2\ell \, \chi _{\omega \ell }  \, P_{i\omega-1/2}^{-3/2-\ell}(\cos\rho)P_{i\omega-1/2}^{-1/2-\ell}(\cos\rho) \sin 2\rho 
  \nonumber \\ & \qquad
     - \left[  2\ell (2\ell + 1 ) +  2(1-\ell ^{2}-5\omega ^{2} ) \sin^{2}\rho \right] [P_{i\omega-1/2}^{-1/2-\ell}(\cos\rho) ]^{2}
      \Bigg\} ,
     \label{eq:SETR_vac11}
\\
      \langle {\hat {T}}_\rho^\rho \rangle _{\text{R},0}^\zeta & =\frac{\cos\rho\cot^3\rho}{192L^4\pi^2} \sum_{\ell=0}^\infty\, \int_{\omega=-\infty}^{\infty} \,D_{\omega \ell}^\zeta \Bigg\{ 6\,\chi_{\omega \ell}^2\, [P_{i\omega-1/2}^{-3/2-\ell} (\cos\rho)]^2 \,\sin^2\rho
           \nonumber \\ & \qquad
            +2(3\ell + 2) \,\chi_{\omega \ell}\,P_{i\omega-1/2}^{-3/2-\ell}(\cos\rho)P_{i\omega-1/2}^{-1/2-\ell}(\cos\rho) \sin2\rho 
            \nonumber \\ & \qquad
     + \left[ 2\ell -2 (1+4\ell + 3\ell ^{2}+3\omega ^{2} ) \sin ^{2} \rho  \right] [P_{i\omega-1/2}^{-1/2-\ell}(\cos\rho) ] ^{2}
        \Bigg\} ,
         \label{eq:SETR_vac22}
\\
     \langle {\hat {T}}_\theta^\theta \rangle _{\text{R},0}^\zeta & =\frac{\cos\rho\cot^3\rho}{192L^4\pi^2}\sum_{\ell=0}^\infty\, \int_{\omega=-\infty}^{\infty} \,D_{\omega \ell}^\zeta \Bigg\{-2\,\chi_{\omega \ell}^2\,[P_{i\omega-1/2}^{-3/2-\ell}(\cos\rho)]^2\sin^2\rho 
                    \nonumber \\ & \qquad
     -2(\ell+1)\,\chi_{\omega \ell}\,P_{i\omega-1/2}^{-3/2-\ell}(\cos\rho)P_{i\omega-1/2}^{-1/2-\ell} (\cos\rho)\sin2\rho 
               \nonumber \\ & \qquad
     + 2 \left[ \ell ^{2} + (1+2\ell + \ell ^{2}-\omega ^{2}) \sin ^{2} \rho  \right]
     [P_{i\omega-1/2}^{-1/2-\ell}(\cos\rho)]^{2}
\Bigg \} ,
     \label{eq:SETR_vac33}
  \end{align}
 with $\langle {\hat {T}}_\phi^\phi \rangle _{\text{R},0}^\zeta=\langle {\hat {T}}_\theta^\theta \rangle _{\text{R},0}^\zeta$, where $D_{\omega \ell}^\zeta$ is given by \eqref{eq:Dconstants}, and $\chi_{\omega \ell}$ is  
  \begin{align}
     \chi_{\omega \ell}&=1+2\ell+\ell^2+\omega^2 .
     \label{eq:UpChi}
  \end{align} 
  For the corresponding expressions for the t.e.v.s we replace $\omega$ with $n\kappa$ and change the integral to a sum to obtain the following nonzero components
   \begin{align}
      \langle {\hat {T}}_\tau^\tau \rangle _{\text{R},\beta}^\zeta & =\frac{\kappa \cos\rho\cot^3\rho}{192L^4\pi^2} \sum_{\ell=0}^\infty\, \sum_{n=-\infty}^{\infty} \, D_{n \ell}^\zeta \Bigg\{ -2\,\chi_{n\ell}^2\, [P_{in\kappa-1/2}^{-3/2-\ell} (\cos\rho)]^2 \,\sin^2\rho
  \nonumber \\ & \qquad
  - 2\ell \, \chi _{n \ell }  \, P_{in\kappa -1/2}^{-3/2-\ell}(\cos\rho)P_{in\kappa -1/2}^{-1/2-\ell}(\cos\rho) \sin 2\rho 
  \nonumber \\ & \qquad
     - \left[  2\ell (2\ell + 1 ) +  2(1-\ell ^{2}-5n^{2}\kappa^{2} ) \sin^{2}\rho \right] [P_{in\kappa-1/2}^{-1/2-\ell}(\cos\rho) ]^{2}
     \Bigg\} ,
     \label{eq:SETR_therm11}
\\
      \langle {\hat {T}}_\rho^\rho \rangle _{\text{R},\beta}^\zeta & =\frac{\kappa\cos\rho\cot^3\rho}{192L^4\pi^2} \sum_{\ell=0}^\infty\, \sum_{n=-\infty}^{\infty} \,D_{n\ell}^\zeta \Bigg\{ 6\,\chi_{n\ell}^2\, [P_{in\kappa-1/2}^{-3/2-\ell} (\cos\rho)]^2 \,\sin^2\rho
      \nonumber \\ & \qquad
             +2(3\ell + 2) \,\chi_{n\ell}\,P_{in\kappa-1/2}^{-3/2-\ell}(\cos\rho)P_{in\kappa-1/2}^{-1/2-\ell}(\cos\rho) \sin2\rho 
            \nonumber \\ & \qquad
     + \left[ 2\ell -2 (1+4\ell + 3\ell ^{2}+3n^{2}\kappa^{2} ) \sin ^{2} \rho  \right] [P_{in\kappa-1/2}^{-1/2-\ell}(\cos\rho) ] ^{2}
     \Bigg\} ,
         \label{eq:SETR_therm22}
\\
     \langle {\hat {T}}_\theta^\theta \rangle _{\text{R},\beta}^\zeta & =\frac{\kappa\cos\rho\cot^3\rho}{192L^4\pi^2}\sum_{\ell=0}^\infty\, \sum_{n=-\infty}^{\infty} \,D_{n\ell}^\zeta \Bigg\{-2\,\chi_{n\ell}^2\,[P_{in\kappa-1/2}^{-3/2-\ell}(\cos\rho)]^2\sin^2\rho 
                    \nonumber \\ & \qquad
     -2(\ell+1)\,\chi_{n \ell}\,P_{in\kappa-1/2}^{-3/2-\ell}(\cos\rho)P_{in\kappa-1/2}^{-1/2-\ell} (\cos\rho)\sin2\rho 
               \nonumber \\ & \qquad
     + 2 \left[ \ell ^{2} + (1+2\ell + \ell ^{2}-n^{2}\kappa^{2}) \sin ^{2} \rho  \right]
     [P_{in\kappa-1/2}^{-1/2-\ell}(\cos\rho)]^{2}
\Bigg\},
     \label{eq:SETR_therm33}
  \end{align} }
and $\langle {\hat {T}}_\phi^\phi \rangle _{\text{R},\beta }^\zeta=\langle {\hat {T}}_\theta^\theta \rangle _{\text{R},\beta }^\zeta$.
The $ D_{n \ell}^\zeta$ and $\chi_{n\ell}$ terms  are obtained from  the corresponding terms in~(\ref{eq:Dconstants}, \ref{eq:UpChi}) by replacing $\omega$ with $n \kappa$.
It is straightforward to check that the contributions to the RSET expectation values arising from the last term in (\ref{eq:SETexp_ren}) have vanishing trace.
  
  The  v.e.v.s and t.e.v.s of the RSET are calculated numerically using {\small MATHEMATICA}. 
  The sums in \eqref{eq:SETexpDN} converge extremely rapidly and are straightforward to compute. 
  The remaining contributions (\ref{eq:SETR_vac11}--\ref{eq:SETR_therm33}) which arise when we impose Robin boundary conditions involve either a double infinite summation (for the t.e.v.s) or an integral and a summation (for the v.e.v.s). 
  For the v.e.v.s, we performed the integral over $\omega $ first before summing over $\ell $.
  For fixed $\ell $, the integral over $\omega $ is rapidly convergent, and we integrated over the interval $|\omega | \le 100$.
  For the t.e.v.s, the sum over $n$ again converges rapidly, and we summed over $n$ with a magnitude of less than or equal to $50$.

\begin{figure}[htbp]
     \begin{center}
        \includegraphics[width=0.9\textwidth]{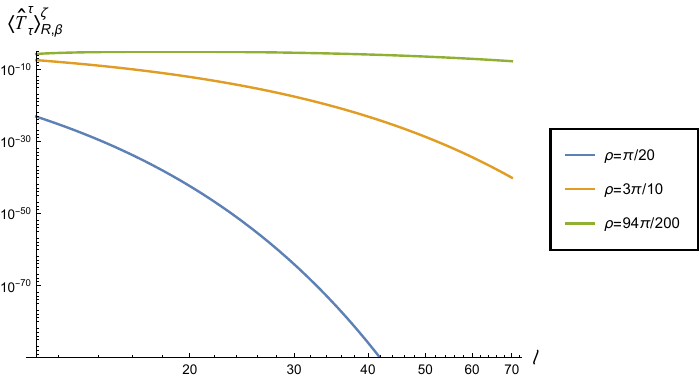} 
               \caption{Log-log plot of the $\ell $-summand in  $\langle {\hat {T}}_\tau^\tau \rangle _{\text{R},\beta}^\zeta $ (\ref{eq:SETR_therm11})  as a function of $\ell $ for a selection of values of the radial coordinate $\rho$. With  $\zeta = 3\pi/5$ and $\kappa=1/2$, we have performed the sum over $|n| \le 50$. 
               It can be seen that, as $\rho $ increases, the $\ell $-summand decreases at a much slower rate with increasing $\ell$, resulting in a sum over $\ell $ which converges more slowly for larger values of $\rho $.}
        \label{fig:therm_lsum_conv}
      \end{center}
\end{figure}

As was found in the computation of the VP \cite{Morley:2021}, the sum over $\ell$ exhibits nonuniform convergence with respect to the radial coordinate, $\rho$, converging more quickly nearer the origin and much slower as the space-time boundary is approached (see Figure~\ref{fig:therm_lsum_conv}).
For the v.e.v.s, we summed over $0\le \ell \le 100$ and for the t.e.v.s $0\le \ell \le 80$.
We used a smaller range of values of $n$ and $\ell $ for the t.e.v.s compared to the v.e.v.s due to increased computation time required for the function evaluations. 
  
We estimated the errors in truncating the sums and integrals as follows. 
For the v.e.v.s, for a selection of values of the Robin parameter $\zeta $ and radial coordinate $\rho $, we compared our results obtained by integrating over $|\omega | \le 100 $ and summing over $0\le \ell \le 100$ with those found from increasing the maximum values of the magnitudes of $\omega $ and $\ell $ to $170$.
For example, for  $\zeta = 3\pi/10$ and  $\rho=94\pi/200$, by this method we estimate the relative error in \eqref{eq:SETR_vac11} to be of order $10^{-2}$.
The relative error was much smaller further away from the space-time boundary, and is estimated to be of order $10^{-18}$ at $\rho=3\pi/10$ and $\rho=\pi/20$.
However, the contributions to the RSET in (\ref{eq:SETR_vac11}-\ref{eq:SETR_vac33}) contribute only a small proportion of the overall value. 
For example, the value of \eqref{eq:SETR_vac11} at $\rho=94\pi/200$ and $\zeta=3\pi/10$ as a fraction of the total v.e.v. of $\langle {\hat {T}}_{\tau }^{\tau}\rangle^\zeta _0$ was $\sim 5 \times 10^{-4}$, meaning that the errors in the numerical calculations of the contributions (\ref{eq:SETR_vac11}-\ref{eq:SETR_vac33}) are much less significant in the final results.
The same holds for the t.e.v.s.
 We employed the same method to estimate the relative errors in the t.e.v.s, and find that the errors depend strongly on both the temperature and the radial coordinate $\rho $. 
 For $\kappa=1/2$ and $\zeta=\pi/10$, the relative  error in the numerical computation of \eqref{eq:SETR_therm11} at $\rho= 80\pi/200$, for example, was $\sim 3.5 \times 10^{-5}$. As a result of increasing errors encountered close to the boundary, the numerical calculation of \eqref{eq:SETR_therm11}, for $\kappa =1/2$ was performed up to $\rho=85\pi/200$ only.  The relative errors near the space-time boundary improved somewhat with increasing $\kappa$. For $\rho=90\pi/200$, for instance, the relative errors in \eqref{eq:SETR_therm11}  were $\sim 2 \times 10^{-10}$ and $\sim 5 \times 10^{-11}$  for $\kappa=2$ and $\kappa= 2\pi$ respectively.

\section{Vacuum expectation value of the RSET with Robin boundary conditions}
\label{sec:vev with Robin}

In the low temperature limit ($\beta \to \infty$), the v.e.v.~of the RSET, $\langle\hat{T}_{\mu\nu}\rangle_{0}^{{\rm {D/N}}}$, derived from  \eqref{eq:SETexpDN} reduces to~\cite{Allen:1986} 
 \begin{equation}
    \langle\hat{T}_{\mu\nu}\rangle_{0}^{{\rm {D/N}}} = - \frac{1}{960 \pi ^{2}L^{4}} g_{\mu \nu }
    \label{eq:trace}
 \end{equation}
 in agreement with the calculation in \cite{Kent:2014nya} for a scalar field with Dirichlet boundary conditions.
 In particular, the v.e.v.~(\ref{eq:trace}) is identical for Dirichlet and Neumann boundary conditions. 
 This does not occur for the VP (where the v.e.v.s for Dirichlet and Neumann boundary conditions are different), and can be understood as follows.
 For both Dirichlet and Neumann boundary conditions, the vacuum state is maximally symmetric, and therefore the RSET will be a constant multiple of the metric, $\langle\hat{T}_{\mu\nu}\rangle_{0}^{{\rm {D/N}}}=\alpha g_{\mu \nu }$ for some constant $\alpha $. 
 Taking the trace, $\alpha = \langle\hat{T}_{\mu}^{\mu}\rangle_{0}^{{\rm {D/N}}}/4$.
 For a massless, conformally coupled scalar field, the trace $\langle\hat{T}_{\mu}^{\mu}\rangle_{0}^{{\rm {D/N}}}$ is fixed to be the trace anomaly, which, on four-dimensional adS is \cite{Kent:2014nya}
 \begin{equation}
    \langle\hat{T}_{\mu}^{\mu}\rangle = -\frac{1}{240\pi ^{2}L^{4}}.
    \label{trace_anom}
 \end{equation}
 Therefore, for a massless, conformally coupled scalar field, the RSET for a maximally symmetric state is entirely determined by the trace anomaly and is independent of any boundary conditions applied.
     
\begin{figure}[htbp]
     \begin{center}
     \begin{tabular}{cc}
        \includegraphics[width=0.40\textwidth]{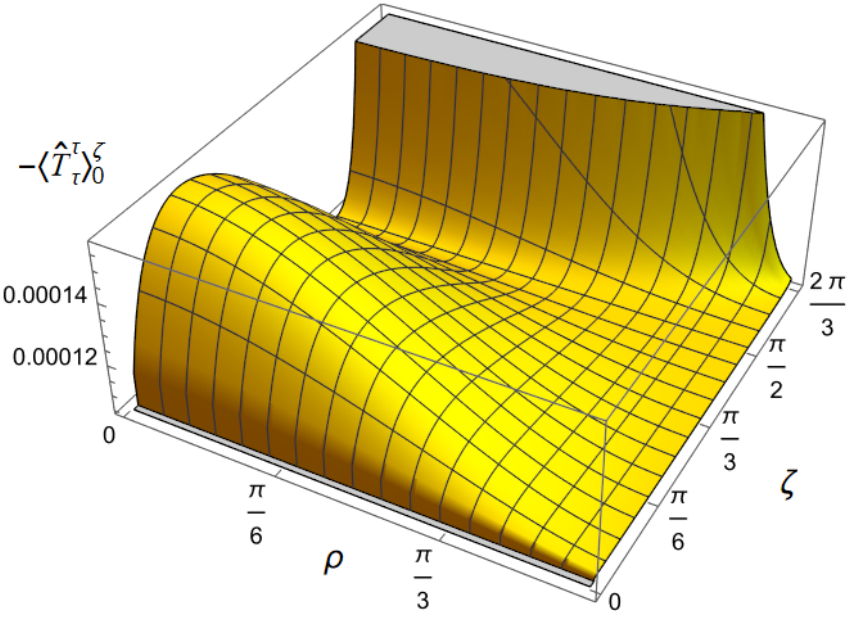} & 
        \includegraphics[width=0.50\textwidth]{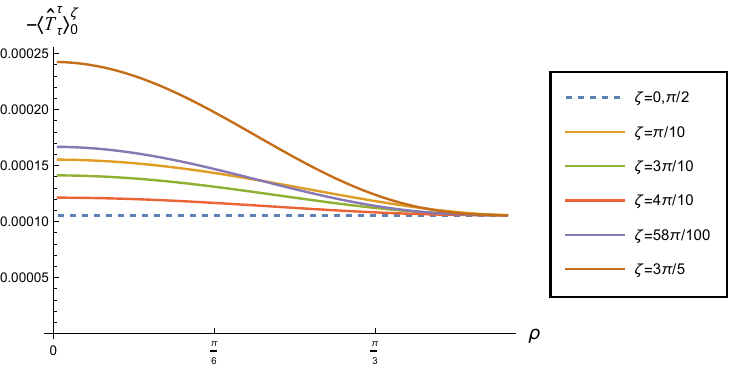}
        \\[0.3cm]
       \includegraphics[width=0.40\textwidth]{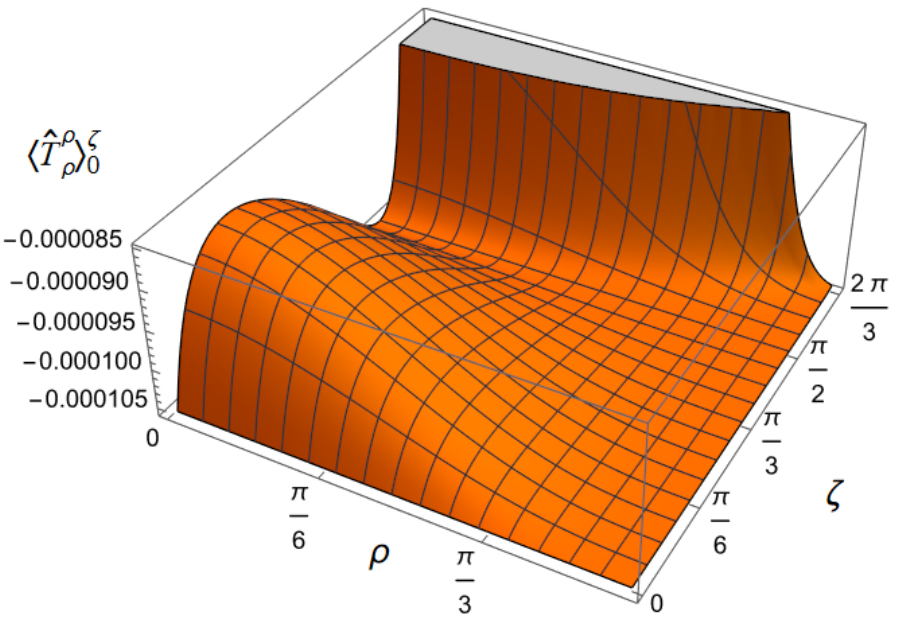} &
       \includegraphics[width=0.50\textwidth]{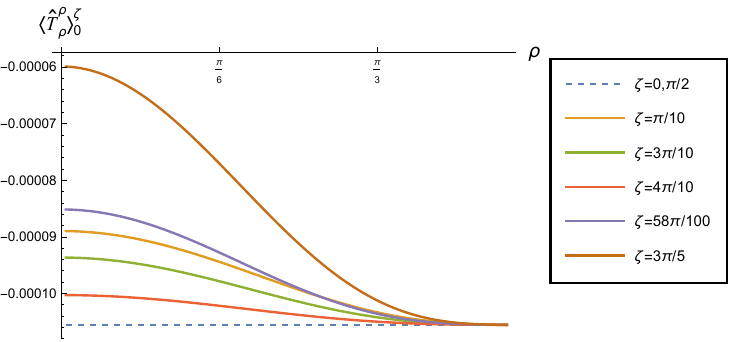}
       \\[0.3cm]
       \includegraphics[width=0.40\textwidth]{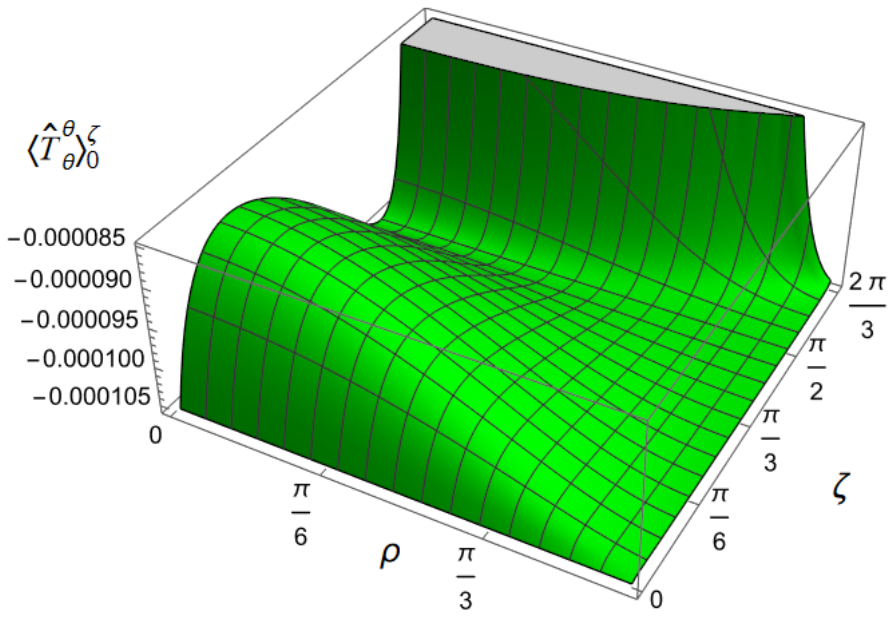} &
       \includegraphics[width=0.50\textwidth]{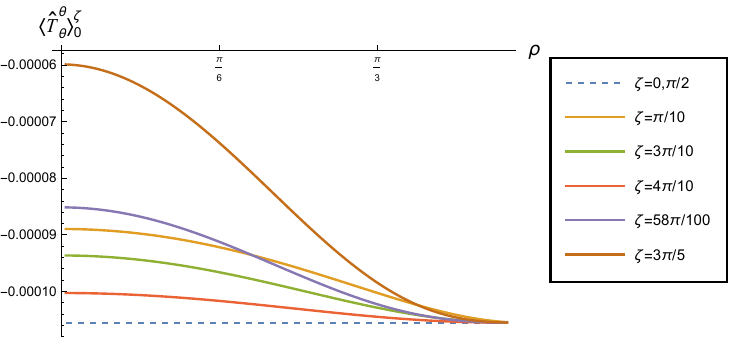}
       \end{tabular}
       \caption{V.e.v.s of the RSET with Robin boundary conditions. The top row shows the energy density $-\langle \hat{T}_\tau^\tau \rangle_{0}^{\zeta }$, the middle row $\langle \hat{T}_\rho^\rho \rangle_{0}^{\zeta }$ and the bottom row $\langle \hat{T}_\theta ^\theta \rangle_{0}^{\zeta }$.   On the left are 3D surface plots showing the variation of the nonzero components of  $\langle {\hat {T}}_\mu^\nu  \rangle_{0}^{\zeta }$ with $\rho$ and $\zeta$. On the right are the nonzero components of $\langle {\hat {T}}_\mu^\nu  \rangle_{0}^{\zeta }$ as functions of $\rho$ for a selection of values of the Robin parameter $\zeta$. Dirichlet and Neumann boundary conditions are shown as dashed lines.}
       \label{fig:vac_SET_Robin}
          \end{center}
   \end{figure}
   \begin{figure}[htbp]
\begin{center}
     \begin{tabular}{cc}
        \includegraphics[width=0.4\textwidth]{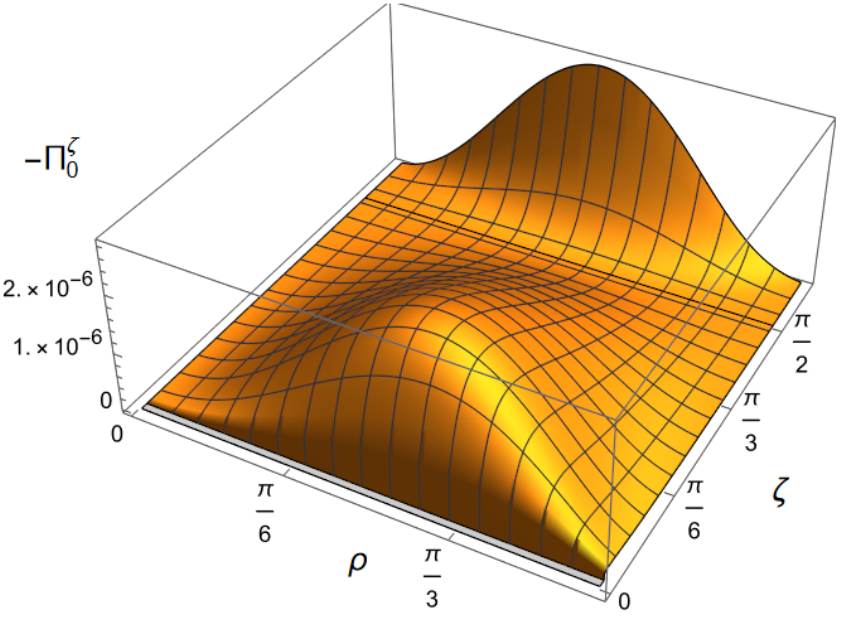} 
        &
       \includegraphics[width=0.54\textwidth]{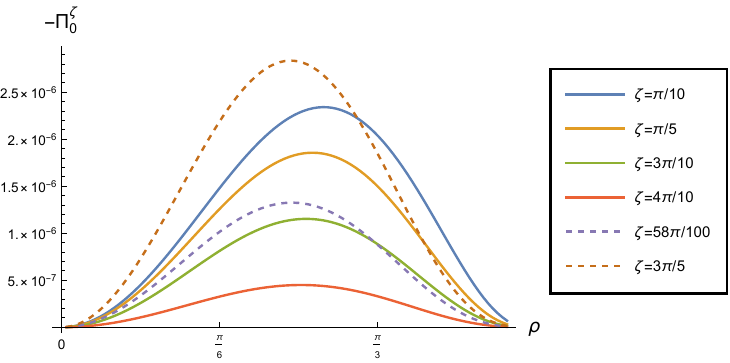}
       \end{tabular}
       \caption{Vacuum pressure deviator $-\Pi_0^\zeta$ (\ref{eq:Landau})  with Robin boundary conditions. On the left is a 3D surface plot showing the variation of $-\Pi_0^\zeta$ with $\rho$ and $\zeta$. On the right is $-\Pi_0^\zeta$ as a function of $\rho$ for a selection of values of the Robin parameter $\zeta$. Robin parameters $\zeta > \pi/2$ are shown with dotted curves. The pressure deviator vanishes identically when Dirichlet or Neumann boundary conditions are applied and is not plotted in these cases.}
       \label{fig:press_dev}
          \end{center}
   \end{figure}
 While the v.e.v of the RSET with either Dirichlet or Neumann boundary respects the maximum symmetry of the underlying space-time, this is not the case when Robin boundary conditions are applied, 
as can be seen in  Figure~\ref{fig:vac_SET_Robin}.   
For all values of $\zeta \ne 0,\pi/2$  each component of the RSET varies with the radial coordinate $\rho $.  The energy density $-\langle \hat{T}_\tau^\tau \rangle _0^\zeta$ is positive throughout the space-time, reaching the common vacuum Dirichlet/Neumann value at the space-time boundary. This is in contrast to the findings in~\cite{Barroso:2019cwp} where the energy density is negative on most of the space-time and only becomes positive as the boundary is reached. This is due to the application of Robin boundary conditions to only a subset of the modes in~\cite{Barroso:2019cwp}.    The other components of the RSET take the same constant values when $\zeta =0$ or $\pi /2$ and Dirichlet or Neumann boundary conditions are applied. 
The quantities plotted in Figure~\ref{fig:vac_SET_Robin} are greatest at the space-time origin and converge to the Dirichlet/Neumann value as the space-time boundary is reached ($\rho \to \pi/2$). 
Their values at the space-time origin increase as $\zeta $ increases from zero, attain a maximum at some value of $\zeta \in (0, \pi/2)$ and then decrease as $\zeta $ approaches $\pi /2$. 
As $\zeta $ increases above $\pi /2$, these quantities increase rapidly as $\zeta $ approaches $\zeta _{{\text {crit}}}\approx 0.68\pi $. 
As discussed at the end of Section~\ref{sec:Euclid}, for values of $\zeta $ greater than $\zeta _{{\text {crit}}}$, there are classical mode solutions of the scalar field equation which are unstable \cite{Morley:2021}. Due to this classical instability, the semiclassical approximation employed in this paper will break down for $\zeta _{{\text{crit}}}<\zeta < \pi $.

Whilst the v.e.v.~of the $\langle \hat{T}_\rho^\rho \rangle$ and $\langle \hat{T}_\theta^\theta \rangle$  components of the RSET are negative, 
they  are greater (less negative) for Robin boundary conditions than when Dirichlet/Neumann boundary conditions are applied.
However, the variation in the v.e.v.s of the RSET components due to varying the boundary conditions is rather small, at roughly the percent level. 

While it may appear from Figure~\ref{fig:vac_SET_Robin} that the v.e.v.s of the $\langle {\hat {T}}_\rho^\rho  \rangle_0^{\zeta }$ and $\langle {\hat {T}}_\theta^\theta  \rangle_{0}^{\zeta }$ components are the same, there is  in fact a subtle difference. Writing the components of the RSET in the Landau decomposition, analogous to that employed in the thermal state~\cite{Ambrus:2018olh}, gives
\begin{equation}
    \langle {\hat {T}}_\mu^\nu \rangle_{0}^{\zeta }= \text{Diag} \Big\{ -E_{0}^{\zeta },\, P_{0}^{\zeta } +\Pi_{0}^\zeta, P_{0}^{\zeta }-\frac{1}{2}\Pi_{0}^\zeta,P_{0}^{\zeta }-\frac{1}{2}\Pi_{0}^\zeta \Big \}
    \label{eq:Landau}
\end{equation}
where $E_{0}^{\zeta }$ is the energy density, $P_{0}^{\zeta }$ the pressure and  $\Pi^\zeta_{0}$ is the shear stress or pressure deviator \cite{Ambrus:2018olh}. The pressure deviator measures the difference between the RSET of the  quantum scalar field compared with that found if the field were modelled as a classical gas of particles (as it vanishes in the latter case). As the $\langle {\hat {T}}_\theta^\theta  \rangle_{0}^{\zeta }$ component of the RSET is greater than the $\langle {\hat {T}}_\rho^\rho  \rangle_{0}^{\zeta }$, component, in Figure~\ref{fig:press_dev} we show $-\Pi^\zeta_0$ as a function of the radial coordinate $\rho$ for the vacuum state. For both Dirichlet and Neumann boundary conditions, the vacuum pressure deviator is zero (not shown in Figure~\ref{fig:press_dev}). For Robin boundary conditions $\Pi^\zeta_0$  vanishes at both the origin and boundary of the space-time and attains its maximum absolute value between the two.
  
  \section{Thermal expectation value of the RSET with Robin boundary conditions}
\label{sec:tev with Robin} 
\begin{figure}[htbp]
     \begin{center}
     \begin{tabular}{cc}
        \includegraphics[width=0.4\textwidth]{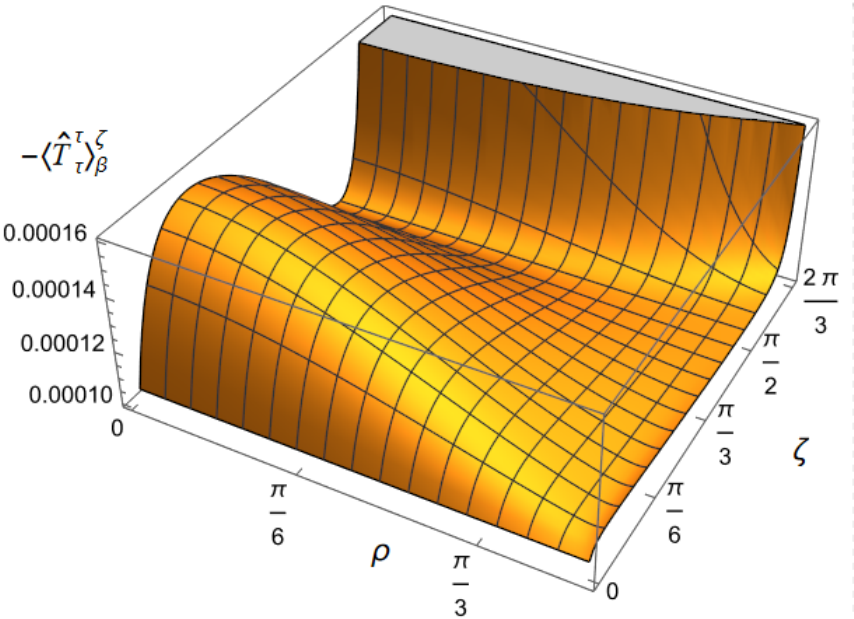} & 
        \includegraphics[width=0.54\textwidth]{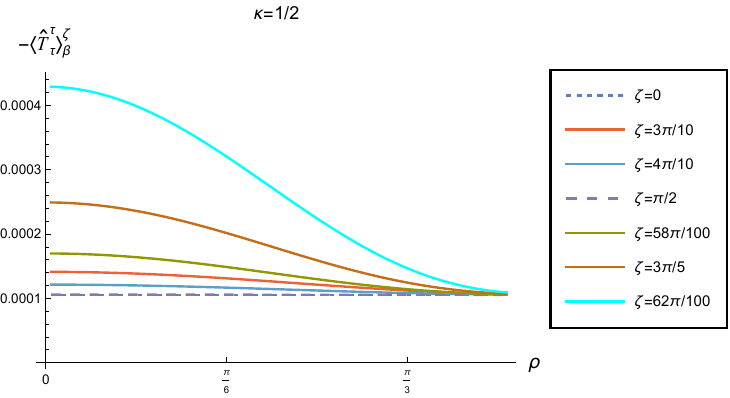}
        \\[0.30cm]
       \includegraphics[width=0.4\textwidth]{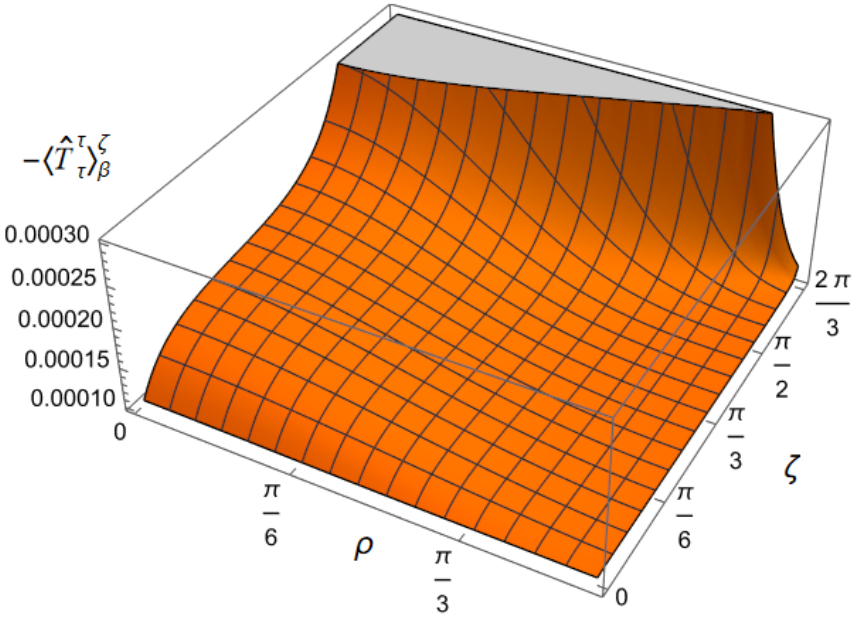} &
       \includegraphics[width=0.54\textwidth]{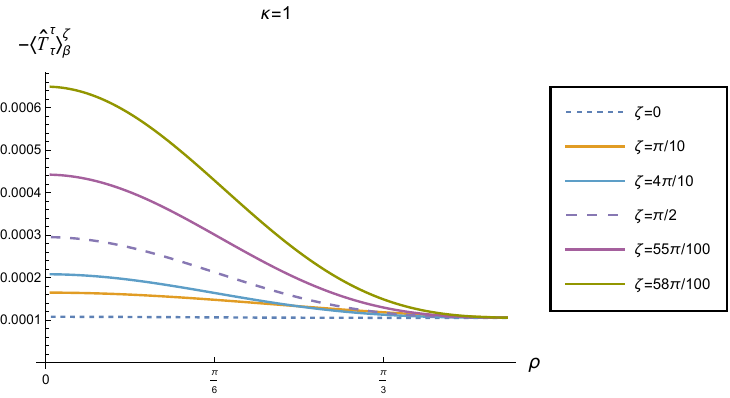}
       \\[0.30cm]
       \includegraphics[width=0.4\textwidth]{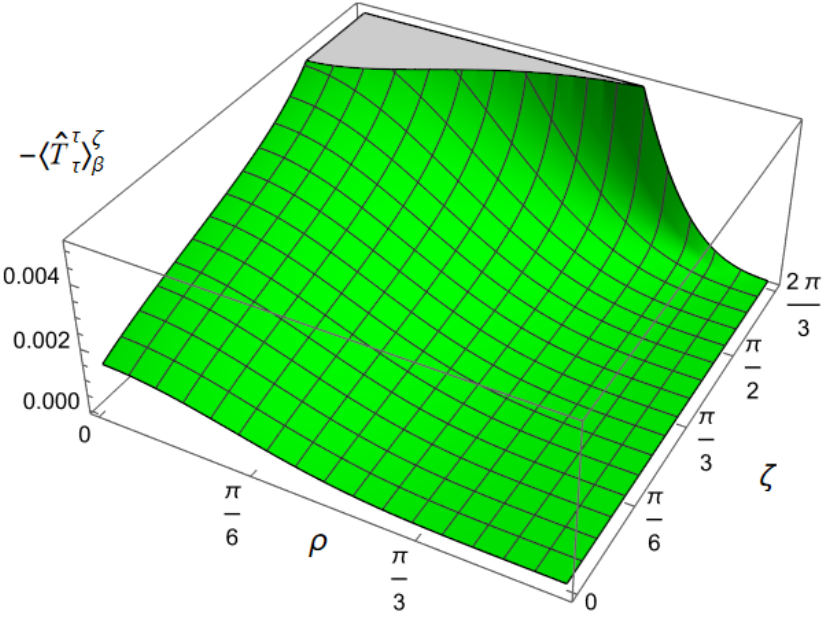} &
       \includegraphics[width=0.54\textwidth]{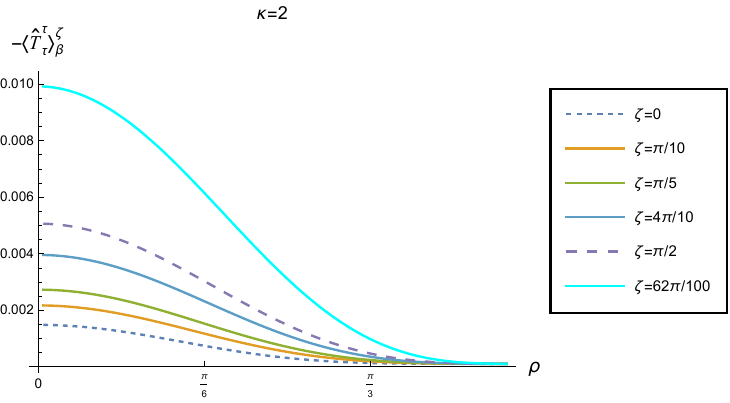}
       \\[0.30cm]
       \includegraphics[width=0.4\textwidth]{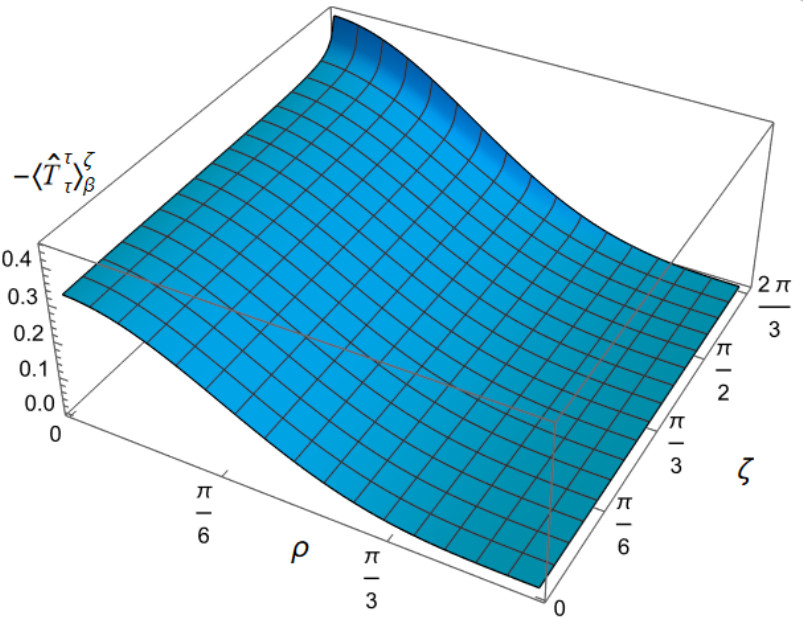} &
       \includegraphics[width=0.54\textwidth]{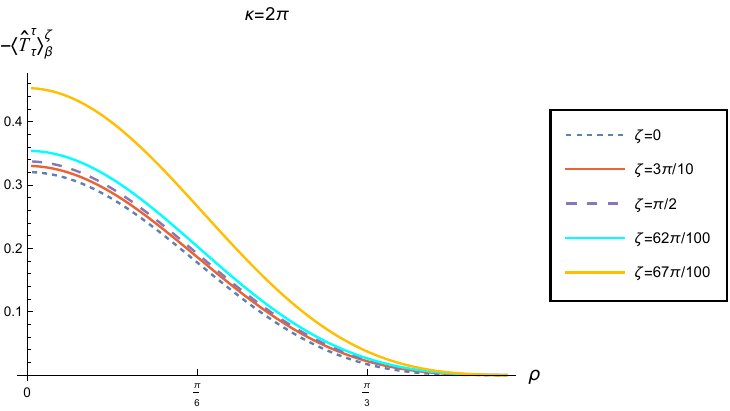}
       \end{tabular}
       \caption{T.e.v.s of the energy density $-\langle {\hat {T}}_\tau^\tau  \rangle_{\beta }^{\zeta }$,  with Robin boundary conditions and a selection of values of $\kappa$ (\ref{eq:kappa}).  On the left are 3D surface plots showing the variation of  $-\langle {\hat {T}}_\tau^\tau  \rangle_{\beta }^{\zeta }$ with $\rho$ and $\zeta$. On the right is $-\langle {\hat {T}}_\tau^\tau  \rangle_{\beta }^{\zeta }$ as a function of $\rho$ for a selection of values of the Robin parameter $\zeta$. 
       Dirichlet and Neumann boundary conditions are shown with dotted lines.}
       \label{fig:therm_SET11_Robin}
          \end{center}
   \end{figure}

  \begin{figure}[htbp]
     \begin{center}
     \begin{tabular}{cc}
        \includegraphics[width=0.4\textwidth]{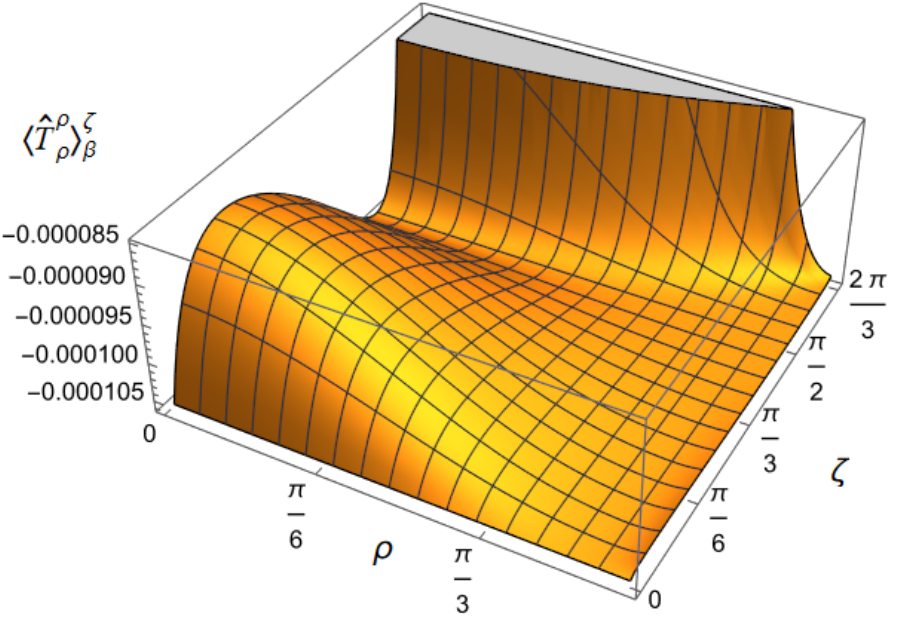} & 
        \includegraphics[width=0.54\textwidth]{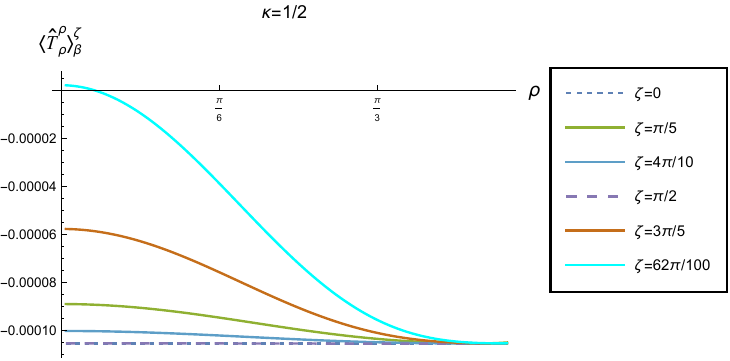}
        \\[0.30cm]
       \includegraphics[width=0.4\textwidth]{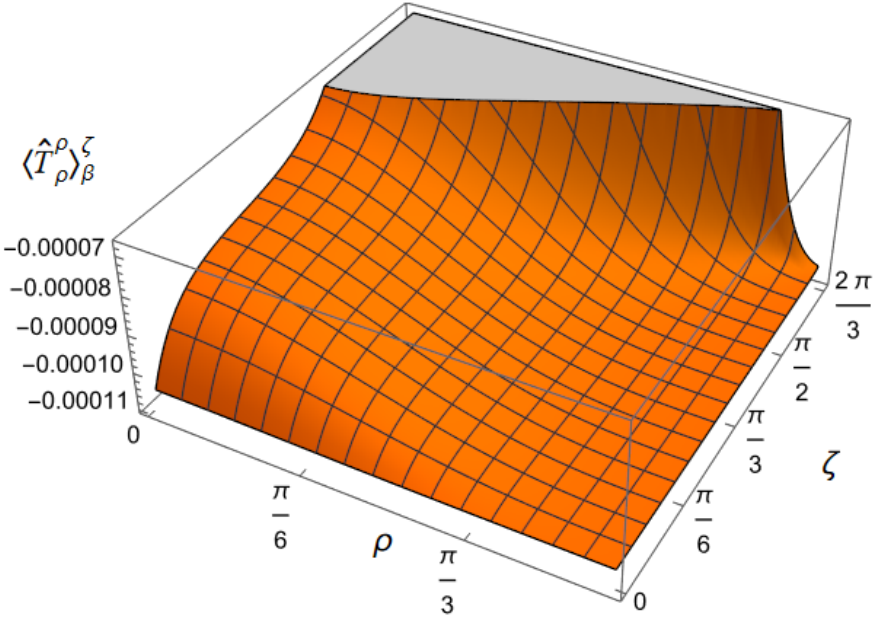} &
       \includegraphics[width=0.54\textwidth]{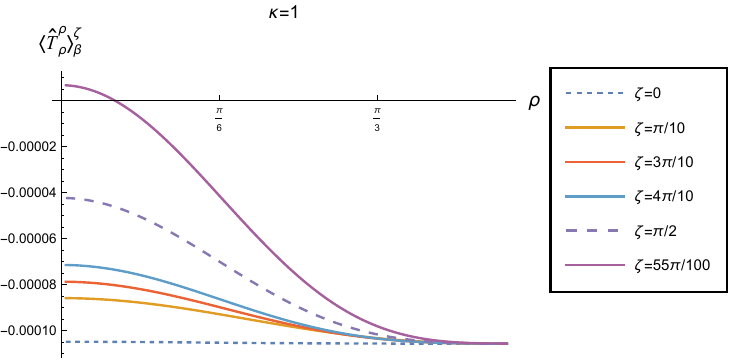}
       \\[0.30cm]
       \includegraphics[width=0.4\textwidth]{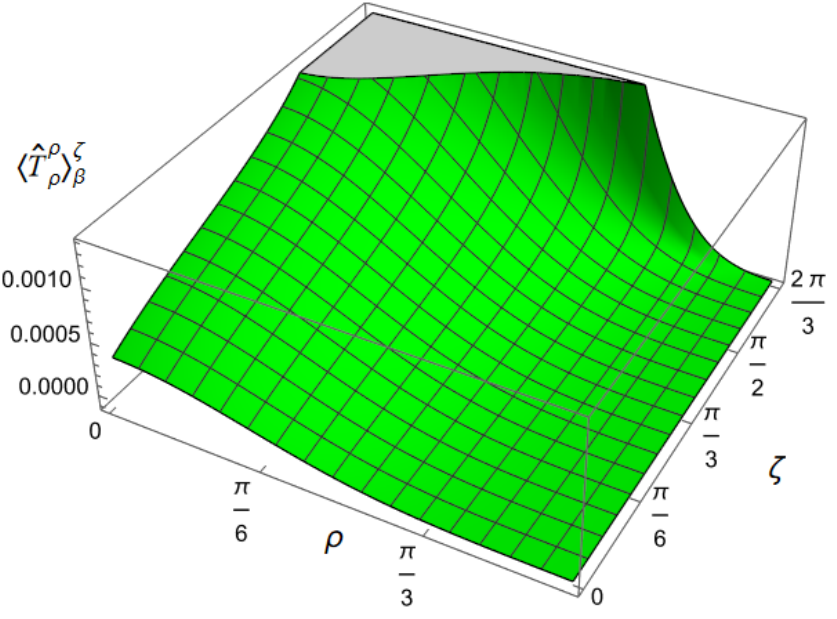} &

       \includegraphics[width=0.54\textwidth]{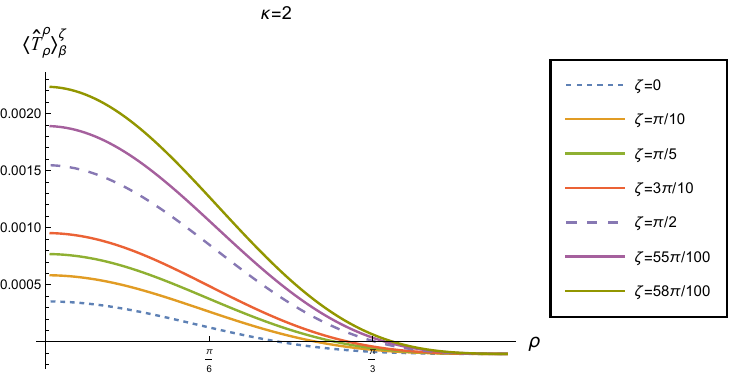}
       \\[0.30cm]
       \includegraphics[width=0.4\textwidth]{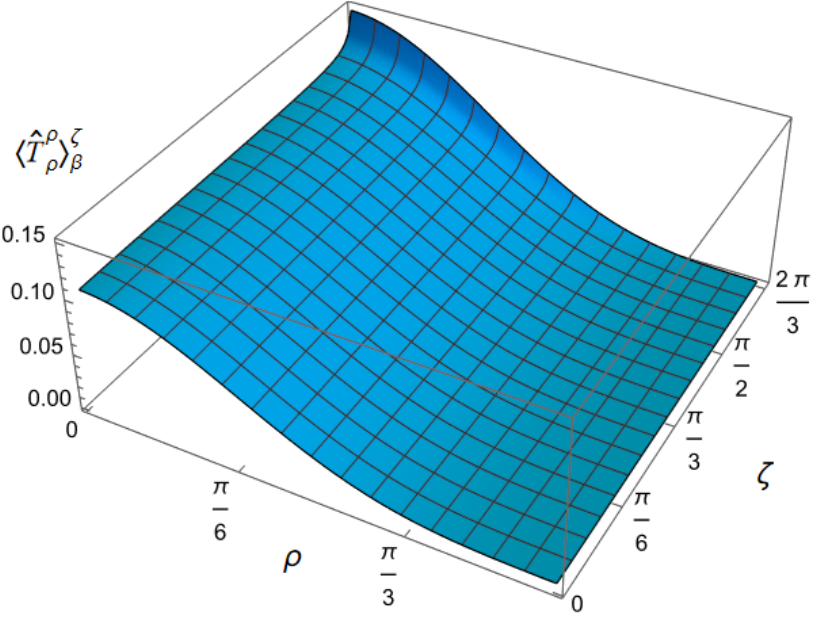} &
       \includegraphics[width=0.54\textwidth]{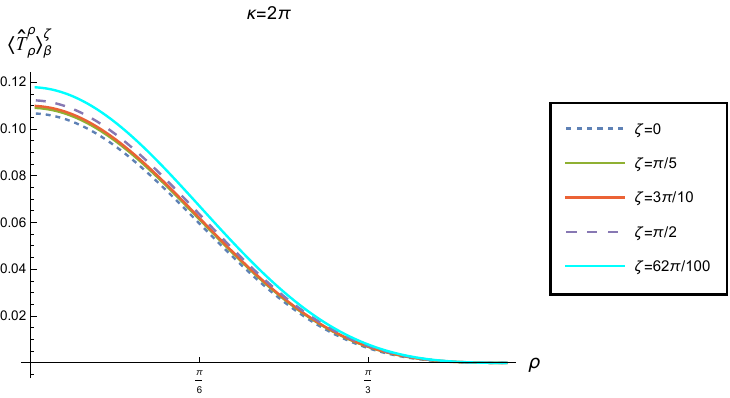}
       \end{tabular}
       \caption{T.e.v.s of the RSET, $\langle \hat {T}_\rho^\rho  \rangle_{\beta }^{\zeta }$,  with Robin boundary conditions and a selection of values of $\kappa$ (\ref{eq:kappa}).  On the left are 3D surface plots showing the variation of  $\langle {\hat {T}}_\rho^\rho  \rangle_{\beta }^{\zeta }$ with $\rho$ and $\zeta$. On the right is $\langle {\hat {T}}_\rho^\rho  \rangle_{\beta }^{\zeta }$ as a function of $\rho$ for a selection of values of the Robin parameter $\zeta$. 
       Dirichlet and Neumann boundary conditions are shown with dotted lines.}
       \label{fig:therm_SET22_Robin}
          \end{center}
   \end{figure}

 \begin{figure}[htbp]
     \begin{center}
     \begin{tabular}{cc}
        \includegraphics[width=0.4\textwidth]{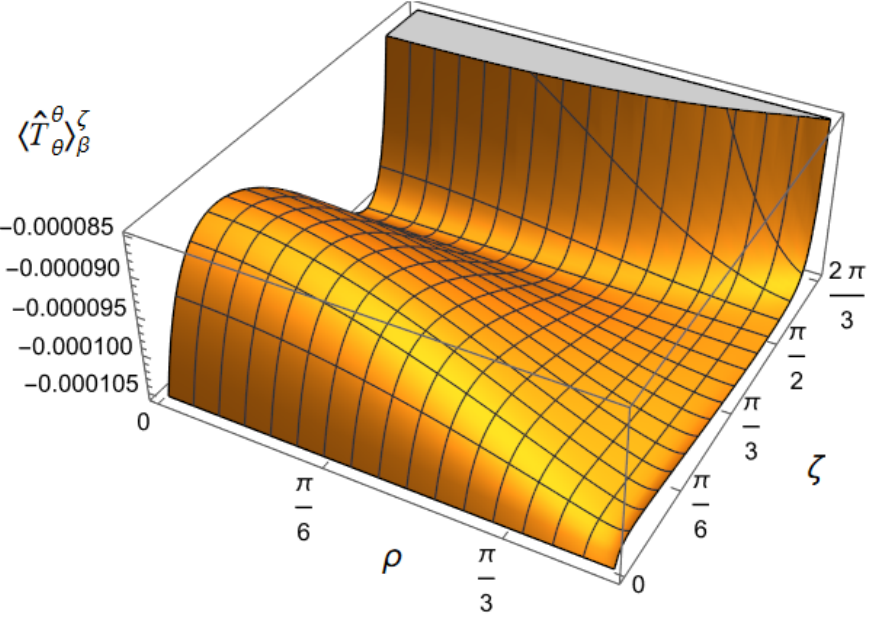} & 
        \includegraphics[width=0.54\textwidth]{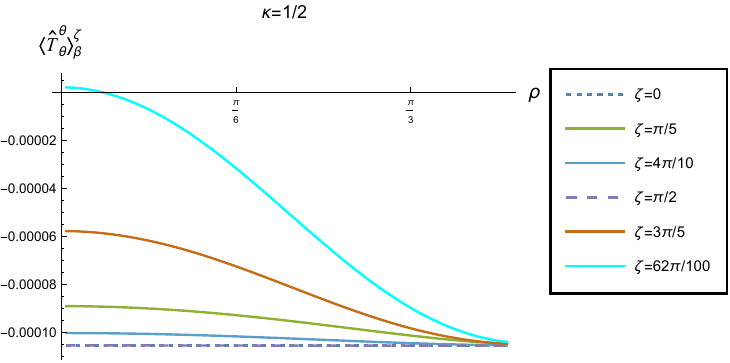}
        \\[0.30cm]
       \includegraphics[width=0.4\textwidth]{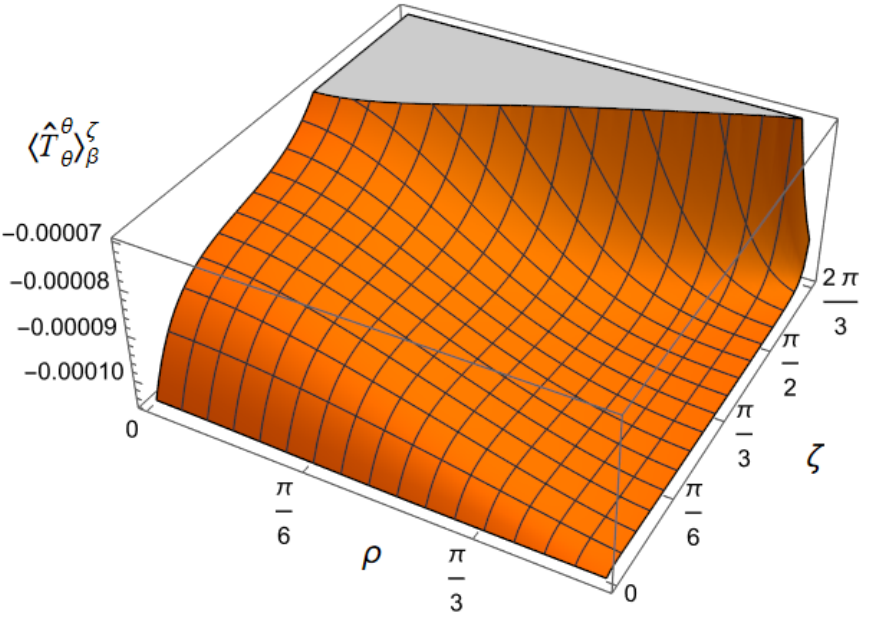} &
       \includegraphics[width=0.54\textwidth]{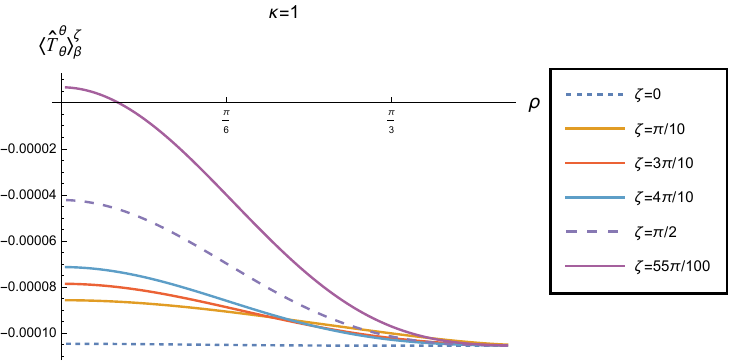}
       \\[0.30cm]
       \includegraphics[width=0.4\textwidth]{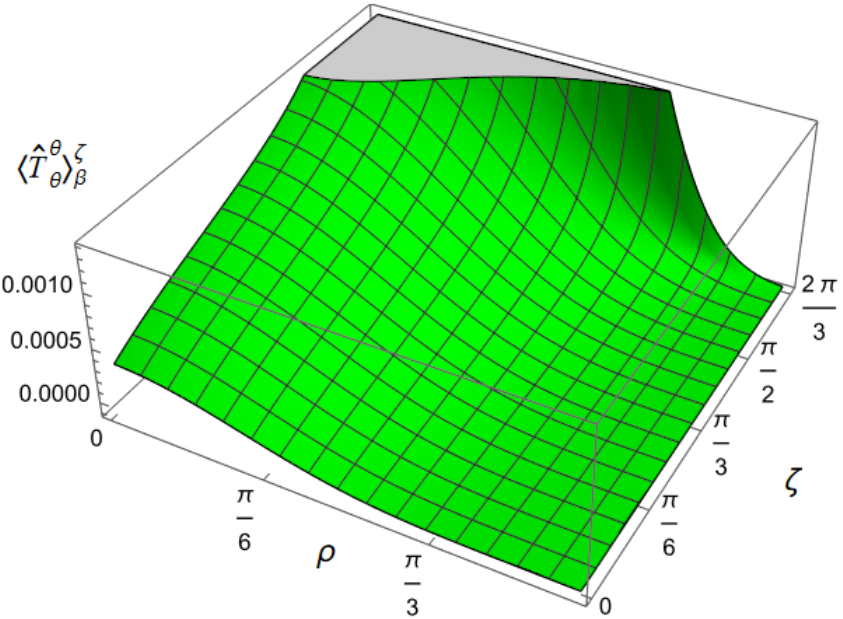} &
       \includegraphics[width=0.54\textwidth]{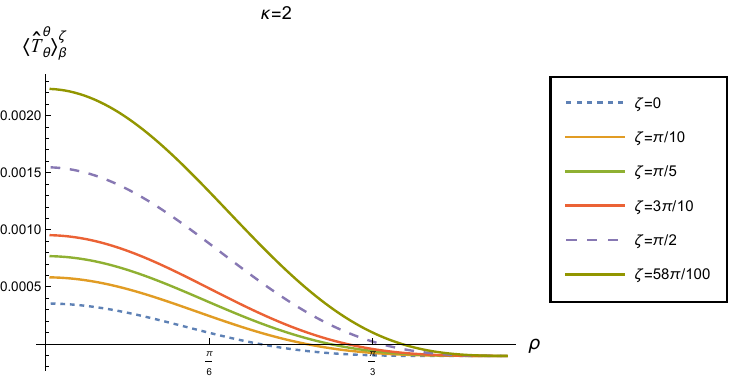}
        \\[0.30cm]
       \includegraphics[width=0.4\textwidth]{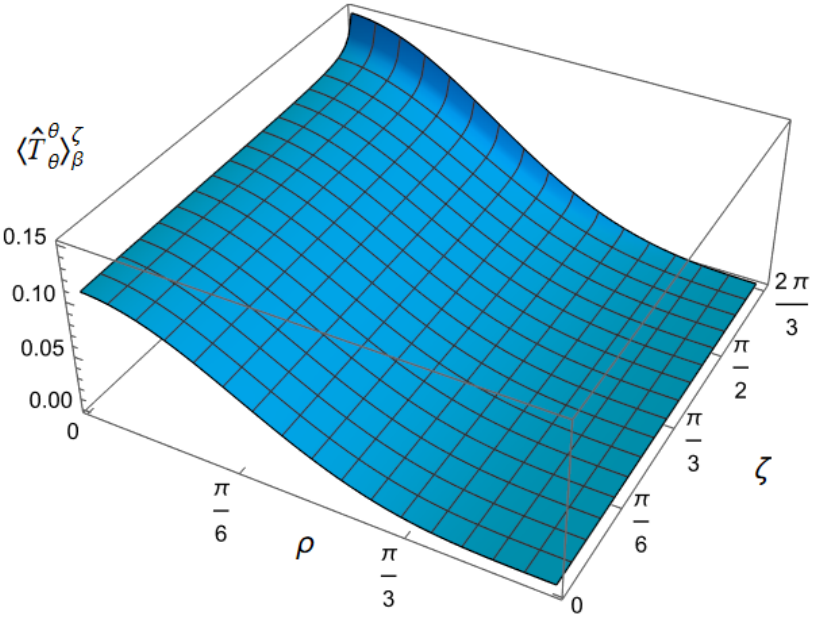} &
       \includegraphics[width=0.54\textwidth]{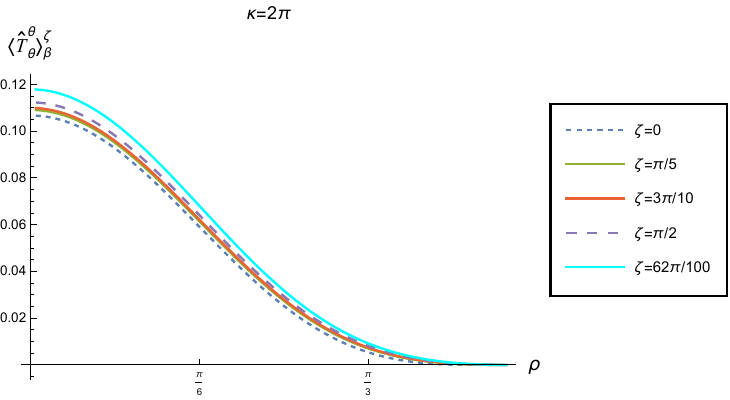}
             \end{tabular}
       \caption{T.e.v.s of the RSET, $\langle {\hat {T}}_\theta^\theta   \rangle_{\beta }^{\zeta }$,  with Robin boundary conditions and a selection of values of $\kappa$ (\ref{eq:kappa}).  On the left are 3D surface plots showing the variation of  $\langle {\hat {T}}_\theta^\theta  \rangle_{\beta }^{\zeta }$ with $\rho$ and $\zeta$. On the right is $\langle {\hat {T}}_\theta^\theta  \rangle_{\beta }^{\zeta }$ as a function of $\rho$ for a selection of values of the Robin parameter $\zeta$. 
       Dirichlet and Neumann boundary conditions are shown with dotted lines.}
       \label{fig:therm_SET33_Robin}
          \end{center}
   \end{figure}

     \begin{figure}[htbp]
     \begin{center}
     \begin{tabular}{cc}
        \includegraphics[width=0.4\textwidth]{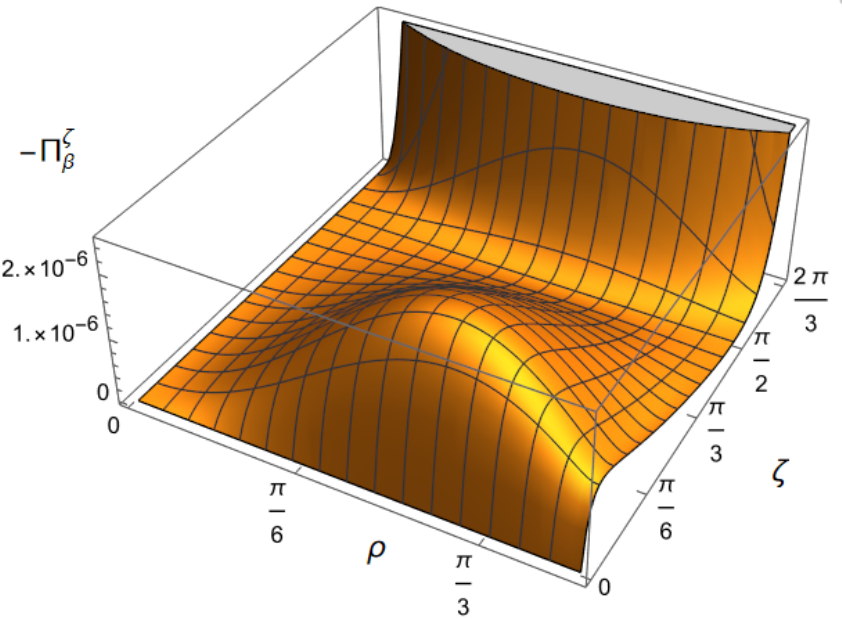} & 
        \includegraphics[width=0.54\textwidth]{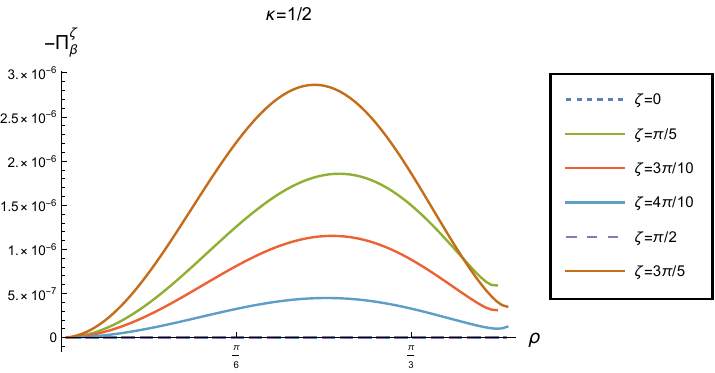}
        \\[0.30cm]
       \includegraphics[width=0.4\textwidth]{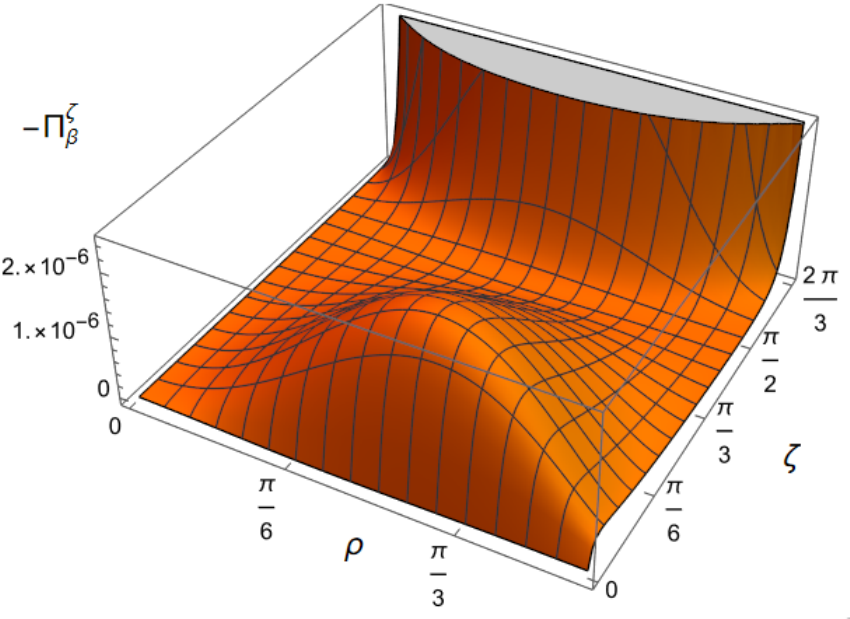} &
       \includegraphics[width=0.54\textwidth]{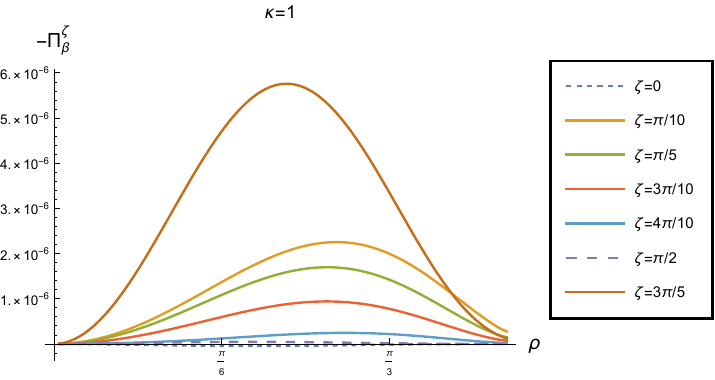}
       \\[0.30cm]
       \includegraphics[width=0.4\textwidth]{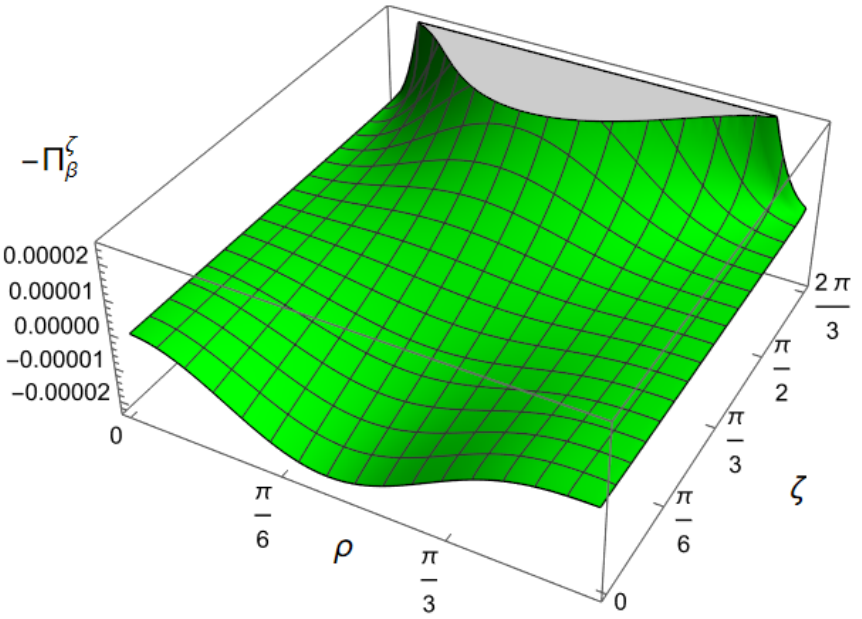} &
       \includegraphics[width=0.54\textwidth]{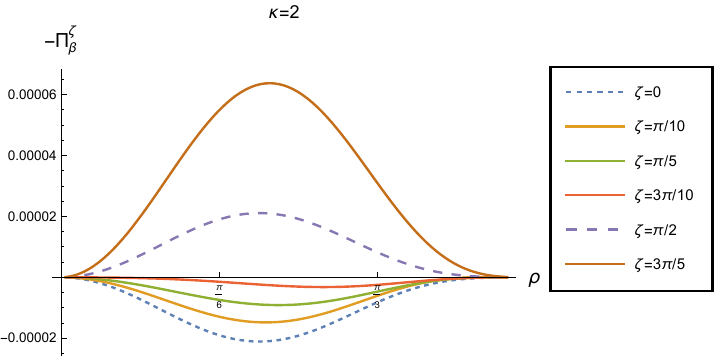}
        \\[0.30cm]
       \includegraphics[width=0.4\textwidth]{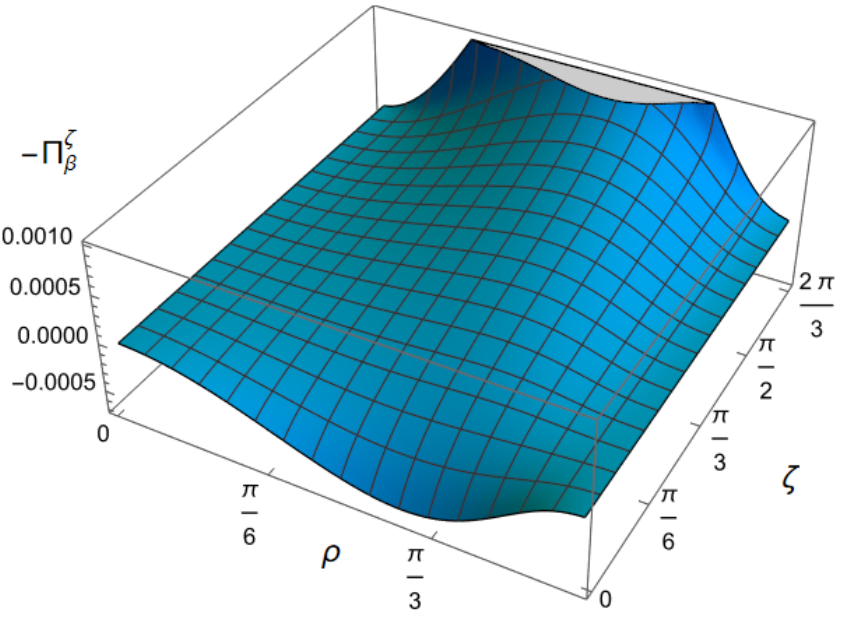} &
       \includegraphics[width=0.54\textwidth]{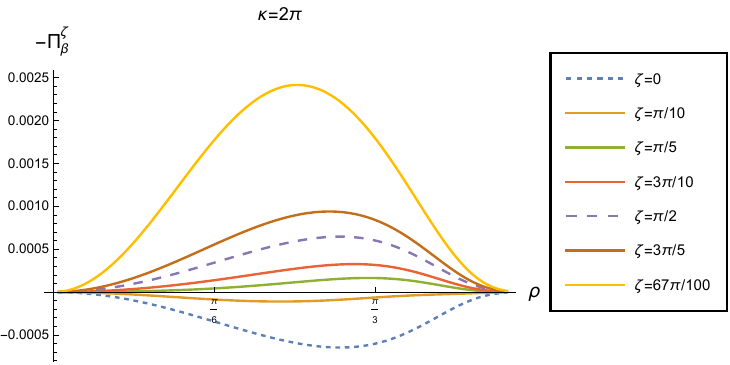}
             \end{tabular}
       \caption{Thermal pressure deviators $-\Pi_\beta^\zeta$ (\ref{eq:Landau})  with Robin boundary conditions and a selection of values of $\kappa $ (\ref{eq:kappa}). On the left are 3D surface plots showing the variation of $-\Pi_\beta ^\zeta$ with $\rho$ and $\zeta$. On the right is $-\Pi_\beta^\zeta$ as a function of $\rho$ for a selection of values of the Robin parameter $\zeta$.  Dirichlet and Neumann boundary conditions are shown with dotted lines.}
       \label{fig:therm_press_dev}
          \end{center}
   \end{figure} 
   
  The t.e.v.s of the nonzero components of the RSET with various values of $\kappa$ (\ref{eq:kappa}) are shown in Figures~\ref{fig:therm_SET11_Robin}-\ref{fig:therm_SET33_Robin} (as in the vacuum state, the $\langle {\hat {T}}_\phi^\phi   \rangle_{\beta }^{\zeta }$ component has the same values as the $\langle {\hat {T}}_\theta^\theta   \rangle_{\beta }^{\zeta }$ component). 
  The nonzero components have very similar behaviour. Unlike the vacuum case, the t.e.v.s with Dirichlet and Neumann boundary conditions, for all nonzero components of the RSET, are no longer constant and vary with the space-time location. 
The difference between the t.e.v.s with Dirichlet and Neumann boundary conditions is a maximum at the space-time origin and decreases with increasing $\rho$. 
The RSET components for these two boundary conditions converge to their common v.e.v.~at the space-time boundary ($\rho \to \pi/2$).  
The absolute difference seen between the nonzero RSET components for the Dirichlet and Neumann boundary conditions increases with increasing $\kappa$ (and hence increasing temperature) and is not clearly discernible at low temperatures in Figures~\ref{fig:therm_SET11_Robin}-\ref{fig:therm_SET33_Robin} due to the scales used.

  The energy density  $-\langle \hat{T}_\tau^\tau \rangle_\beta^\zeta$, is positive throughout the space-time, achieves its maximum value at the space-time origin  and increases with increasing temperature. 
  For all Robin parameters studied, the energy density converges to the common vacuum  Dirichlet/Neumann value at the space-time boundary.   
  For the other nonzero components of the RSET, the t.e.v.s are predominantly negative at low temperature and increase with increasing temperature, becoming positive in a neighbourhood of the space-time origin at $\rho =0$ for sufficiently large $\kappa$.
    They   also achieve their maximum values at the space-time origin, converging to the v.e.v.~\eqref{eq:trace} at the space-time boundary.   

  It can be seen that at low temperature ($\kappa=1/2$), the curves for t.e.v.s with Robin boundary conditions lie outside of the curves corresponding to Dirichlet/Neumann boundary conditions. With increasing temperature, the curves for t.e.v.s with Robin boundary conditions increasingly lie within those for Dirichlet/Neumann boundary conditions and are mostly contained within them for $\kappa=2\pi$.
  As the temperature increases, the variation in the nonzero components of the RSET with varying Robin parameter $\zeta $ becomes much less apparent, as seen in Figures~\ref{fig:therm_SET11_Robin}-\ref{fig:therm_SET33_Robin}.

  As in the vacuum case, we also plot minus the thermal pressure deviator ($-\Pi^\zeta_\beta$),  which is
 the difference between the $\langle {\hat {T}}_\rho^\rho   \rangle_{\beta }^{\zeta }$ and $\langle {\hat {T}}_\theta^\theta   \rangle_{\beta }^{\zeta }$ components of the t.e.v.~of the RSET (see 
 Figure~\ref{fig:therm_press_dev}). The pressure deviator is not only sensitive to the different  Robin boundary conditions (as in the vacuum case) but also to the different  temperatures. Unlike the vacuum case,  the pressure deviator is no longer zero  everywhere for Dirichlet and Neumann boundary conditions (see also \cite{Ambrus:2018olh}, where Dirichlet boundary conditions are applied). 
 For these boundary conditions, the pressure deviator does vanish at the space-time origin and boundary and attains its maximum magnitude between these, this maximum magnitude increasing as the temperature increases. 
 There is a difference in sign with $-\Pi_\beta^\zeta$ being negative for Dirichlet  and positive for  Neumann boundary conditions respectively. 

 For Robin boundary conditions, the profile of the pressure deviator is largely similar to that for Dirichlet or Neumann boundary conditions, vanishing at the origin and space-time boundary and having a maximum magnitude at some $\rho \in (0, \pi/2)$.
 At the higher temperatures we see that $\Pi_\beta^\zeta$ is most positive with Dirichlet boundary conditions ($\zeta=0$) but with increasing Robin parameter, $\zeta$, the pressure deviator becomes increasingly negative.
 As seen in the RSET components, we find that with increasing temperature, the thermal pressure deviators with different Robin boundary conditions are increasingly `contained' within the Dirichlet and Neumann curves.

\section{The RSET at the boundary}
\label{sec:boundary}
The behaviour of the RSET components as the space-time boundary is approached may be understood from the corresponding analysis in \cite{Morley:2021} for the VP. 
Since we are considering a massless, conformally coupled scalar field, we can make a conformal transformation to the Einstein static universe (ESU), containing a time-like surface which is the image of the adS boundary under this mapping.
Using the general construction in \cite{Deutsch:1978sc}, the Green's function for the scalar field on ESU with Robin boundary conditions applied can be written as an asymptotic series in terms of the Green's function on ESU with Neumann boundary conditions applied $G^{\rm {ESU}}_{N}(x,x')$ (see \cite{Deutsch:1978sc,Morley:2021} for more details).
This procedure gives the following asymptotic series for the vacuum Euclidean Green's function on ESU, $G^{\rm {ESU}}_{\zeta , 0}$, with Robin boundary conditions applied:
\begin{equation}
G^{\rm {ESU}}_{\zeta,0 }(x,x') = G^{\rm {ESU}}_{{\rm {N}},0}(x,x') -\frac{1}{L}G_{\zeta ,0}^{(1)}(x,x') \cot \zeta  + \frac{1}{L^{2}}G_{\zeta ,0}^{(2)}(x,x') \cot^{2}\zeta  + \ldots 
\label{eq:series}
\end{equation}
where the first two terms in the series are given by
\begin{align}
G_{\zeta ,0}^{(1)}(x,x') & = \int _{{\mathcal{I}}_{\pi/2}} G^{\rm {ESU}}_{\rm {N}}(x,y)G^{\rm {ESU}}_{{\rm {N}}}(y,x') \, dS ,
\\
G_{\zeta ,0}^{(2)}(x,x') & = \int _{{{\mathcal {I}}}_{\frac{\pi }{2}}} G^{\rm {ESU}}_{{\rm {N}}}(x,y) \left[  \int _{{{\mathcal {I}}}_{\frac{\pi }{2}}} G^{\rm {ESU}}_{{\rm {N}}}(y,z)G^{\rm {ESU}}_{{{\rm {N}}}}(z,x') \, dS  \right] \, dS .
\end{align}
Here ${\mathcal {I}}_{\pi /2}$ is the surface at $\rho =\pi/2$ in ESU, and the integrals are performed over the space-time points $y$, $z$ on this surface in ESU.
Higher-order terms in the series can be found iteratively.
The Green's function on ESU with Neumann boundary conditions applied has a compact closed-form expression \cite{Morley:2021}
\begin{equation}
G_{\rm {N}}^{\rm {ESU}}(x,x') = \frac{1}{8\pi ^{2}L^{2}} \left\{ 
\frac{1}{\cosh \Delta \tau + \cos \Psi} + \frac{1}{\cosh \Delta \tau + \cos \Psi ^{*}}
\right\}
\end{equation}
where $\Delta \tau = \tau - \tau' $ is the separation of the points in the $\tau$-direction, 
\begin{align}
    \Psi & = \arccos \left[ -\cos \rho \cos \rho ' - \cos \gamma \sin \rho \sin \rho ' \right] ,
  \\
    \Psi ^{*} & =  \pi + \arccos \left[ -\cos \rho \cos \rho ' + \cos \gamma \sin \rho \sin \rho ' \right]
\end{align}
and $\gamma $ is the angular separation of the points (\ref{eq:gamma}).
Applying the differential operator ${\mathcal {T}}_{\mu \nu }(x,x')$ to the Green's function (\ref{eq:series}) and bringing the space-time points together gives
\begin{multline}
    \langle\hat{T}_{\mu\nu}\rangle_{\zeta,0}^{\rm{ESU}} = \langle\hat{T}_{\mu\nu}\rangle_{\rm {N},0}^{\rm{ESU}}
    - \frac{\cot \zeta }{L} \lim _{x'\rightarrow x} \left\{ {\mathcal {T}}_{\mu \nu }(x,x') 
    \left[ G_{\zeta ,0}^{(1)}(x,x') \right] \right\}  \\
    + \frac {\cot ^{2}\zeta }{L^{2}} \lim _{x'\rightarrow x} \left\{ {\mathcal {T}}_{\mu \nu }(x,x') 
    \left[ G_{\zeta ,0}^{(2)}(x,x')\right] \right\}
    +\ldots 
    \label{assymp}
    \end{multline}
  We can relate the RSET on ESU to that on adS using~\cite{Birrell:1982ix}
  \begin{equation}
      \langle \hat{T}_\mu^\nu \rangle^{\rm{adS}}_{\zeta,0}=\langle \hat{T}_\mu^\nu \rangle^{\rm{ESU}}_{\zeta,0} \frac{\sqrt{\tilde{g}}}{\sqrt{g}}-\frac{1}{2880\pi^2}\left[ \frac{1}{6}\,{}^{(1)}H_\mu^\nu-{}^{(3)}H_\mu^\nu\right],
      \label{adSESU}
  \end{equation}
  where $\tilde{g}$ and $g$  are the determinants of the metrics on ESU and adS respectively and ${}^{(1)}H_{\mu\nu}$ and ${}^{(3)}H_{\mu\nu}$ are given by~\cite{Birrell:1982ix}
  \begin{align}
      {}^{(1)}H_{\mu\nu}&=2R_{;\mu\nu}-2g_{\mu\nu}\Box R-\frac{1}{2}g_{\mu\nu}R^2+2RR_{\mu\nu}, \\
      {}^{(3)}H_{\mu\nu}&=R_\mu^{\,\rho} R_{\rho\nu}-\frac{2}{3}RR_{\mu\nu}-\frac{1}{2}R_{\rho\sigma}R^{\rho\sigma}g_{\mu\nu}+\frac{1}{4}R^2g_{\mu\nu}.
      \label{eq:H1H3}
  \end{align}
On adS, ${}^{(1)}H_{\mu\nu}$ vanishes identically and ${}^{(3)}H_{\mu\nu} = 3g_{\mu \nu }/L^{2}$.
Using (\ref{assymp}, \ref{adSESU}) we can write 
 \begin{multline}
   \langle \hat{T}_\mu^\nu \rangle^{\rm{adS}}_{\zeta,0}=\langle \hat{T}_\mu^\nu \rangle^{\rm{adS}}_{\rm{N},0}   - \frac{\cot \zeta }{L} \lim _{x'\rightarrow x} \left\{ {\mathcal {T}}_{\mu \nu }(x,x') 
    \left[ G_{\zeta ,0}^{(1)}(x,x') \right] \right\} \cos^4 \rho \\
    + \frac {\cot ^{2}\zeta }{L^{2}} \lim _{x'\rightarrow x} \left\{ {\mathcal {T}}_{\mu \nu }(x,x') 
    \left[ G_{\zeta ,0}^{(2)}(x,x')\right] \right\}\cos^4\rho
    +\ldots 
    \label{eq:finalassymp}
 \end{multline}
 From the analysis in \cite{Deutsch:1978sc}, the RSET on ESU  (\ref{assymp}) can also be expressed as an asymptotic series at an arbitrarily small distance, $\epsilon$, from the boundary at $\rho = \pi /2$ as 
 \begin{equation}
     \langle \hat{T}_{\mu\nu} \rangle^{\rm{ESU}}_{\zeta,0} \sim g^{\alpha'}_{\,\mu} g^{\beta'}_{\,\nu} \left (\epsilon^{-4}\, T^{(4)}_{\alpha'\beta'} + \epsilon^{-3}\,T^{(3)}_{\alpha'\beta'}+ \epsilon^{-2}\,T^{(2)}_{\alpha'\beta'} \right ) + O(\epsilon^{-1}),
     \label{eq:asymp3}
 \end{equation} 
 where $g^{\alpha'}_{\, \mu }(x,x')$ is the bivector of parallel transport between the space-time points $x$ and $x'$. 
 When substituted in (\ref{adSESU}), the leading-order term $\epsilon^{-4}g^{\alpha'}_{\,\mu} g^{\beta'}_{\,\nu} T^{(4)}_{\alpha'\beta'}$, together with the contribution from ${}^{(3)}H_{\mu\nu}$, yields $\langle\hat{T}_{\mu\nu}\rangle_{\rm {N},0}^{\rm{adS}}$ in (\ref{eq:finalassymp}).
The next-to-leading order quantity $\epsilon^{-3}g^{\alpha'}_{\,\mu} g^{\beta'}_{\,\nu} T^{(3)}_{\alpha'\beta'}$, corresponds to the second term in the expansion \eqref{assymp}, and the quantity 
$\epsilon^{-2}g^{\alpha'}_{\,\mu} g^{\beta'}_{\,\nu} T^{(2)}_{\alpha'\beta'}$ to the third term in \eqref{assymp}. 
From~\cite{Deutsch:1978sc}, the next-to-leading order term is given, up to a multiplicative constant, by:
 \begin{equation}
     T^{(3)}_{\mu\nu} \propto \left ( 3 \chi_{\mu\nu}-\chi h_{\mu\nu} \right ),
 \end{equation}
where $\chi_{\mu\nu}=n_{\mu;}{}_\alpha h^\alpha_\nu$ and $n_\mu$ is a unit vector normal to the boundary. The nonzero components of $\chi_{\mu\nu}$ are $\chi_{\theta\theta}=L\cos\rho\sin\rho $  and $\chi_{\phi\phi}=L\cos\rho\sin\rho\sin^2\theta$, giving $\chi=2\cot\rho/L$. As we approach the boundary ($\rho \to \pi/2)$, we have $\chi_{\mu\nu}=\chi =0$, and therefore the second term in \eqref{eq:asymp3} is zero.
We arrive at the same conclusion from a direct computation of the second term in the asymptotic expansion (\ref{assymp}).
Subsequent terms in the expansion are of lower order in $\epsilon$. This means that, as we approach the boundary, $ \langle \hat{T}_\mu^\nu \rangle^{\rm{adS}}_{\zeta,0}=\langle \hat{T}_\mu^\nu \rangle^{\rm{adS}}_{\rm{N},0}$ as shown numerically in Section~\ref{sec:vev with Robin}. 
Similar arguments apply to the t.e.v. of the RSET.

\section{Discussion}
\label{sec:discussion}

In this paper we have determined the v.e.v.s and t.e.v.s of the components of the RSET   for a massless, conformally coupled scalar field propagating on a background four-dimensional global adS space-time. 
We have used Euclidean methods, which give a unique Green's function and avoid the need for an `$i\epsilon$' prescription, rendering the numerical calculations easier than in the corresponding Lorentzian case (see, for example \cite{Barroso:2019cwp}, whose results we have been able to reproduce with our Euclidean methods). 

With mixed indices, the v.e.v.s of the nonzero components of  the RSET are constant when both Dirichlet and Neumann boundary conditions are applied,  respecting the underlying maximal symmetry of the adS space-time. 
Furthermore, the constant is fixed by the trace anomaly  (since we are considering a massless, conformally coupled scalar field), and hence is the same for both Dirichlet and Neumann boundary conditions. 
This common value for the v.e.v. with both Dirichlet and Neumann boundary conditions differs from that seen with the VP  \cite{Morley:2021,Namasivayam:2022,Allen:1986} . 
The maximal symmetry is broken when Robin boundary conditions are applied, and  the v.e.v.s depend on the space-time location. 
However, for all Robin boundary conditions, the v.e.v.s of the nonzero components of the RSET with mixed boundary conditions take the same value on the space-time boundary, namely that for Dirichlet and Neumann boundary conditions.

This symmetry breaking is also seen with the t.e.v.s, even for Dirichlet and Neumann boundary conditions.
The t.e.v.s with either Dirichlet or Neumann boundary conditions are no longer constant and depend on the spatial location, with the maximum difference between them being found at the space-time origin. 
For thermal states, the boundary conditions have a significant effect on the expectation values of all nonzero components of the RSET.
This effect is most apparent near the space-time origin, but becomes diluted with increasing temperature. 
With increasing temperature we find that the t.e.v.s with different Robin boundary conditions are increasingly `contained' with the Dirichlet and Neumann curves, with the difference between all boundary conditions proportionately decreasing with increasing temperature. 
However, for all temperatures and Robin parameters, the t.e.v.s of all nonzero components of the RSET with mixed indices converge at the space-time boundary to the common v.e.v.~found  with Dirichlet and Neumann boundary conditions.
This can be compared with the results for the VP \cite{Morley:2021} where the v.e.v.s and t.e.v.s for all Robin parameters converged to the Neumann result, except when Dirichlet boundary conditions were applied. 
In the case of the RSET,  as both Dirichlet and Neumann boundary conditions result in the same v.e.v.s, in this case, all boundary conditions, including Dirichlet, converge to the same result. 
This supports the conclusion in \cite{Morley:2021,Namasivayam:2022} that Neumann boundary conditions reflect the generic behaviour of the quantum scalar field  at the boundary. 

The value of the VP on the boundary is {\it {a priori}} unconstrained by the renormalization process, whereas the RSET for any maximally symmetric state of a massless, conformally coupled scalar field is completely determined by the trace anomaly. 
This is not the case for scalar fields with mass and/or more general coupling to the space-time curvature, when, even for a maximally symmetric state, the trace of the RSET depends on the constant value of the VP as well as the mass and coupling \cite{Kent:2014nya}.
It would therefore be interesting to compute the RSET for massive or nonconformally-coupled scalar fields, extending the work of \cite{Namasivayam:2022}, which we plan to do in a forthcoming paper.

\begin{acknowledgements}
We thank Axel Polaczek for assistance with the symbolic computation of derivatives of Legendre functions. 
The work of E.W.~is supported by the Lancaster-Manchester-Sheffield Consortium for Fundamental Physics under STFC grant ST/T001038/1.
	This research has also received funding from the European Union's Horizon 2020 research and innovation program under the H2020-MSCA-RISE-2017 Grant No.~FunFiCO-777740.  
 Data supporting this publication can be freely downloaded from the University of Sheffield Research Data Repository at {\url {https://doi.org/10.15131/shef.data.24712308}}, under the terms of the Creative Commons Attribution (CC BY) licence.
 This version of the article has been accepted for publication, after peer review, 
but is not the Version of Record and does not reflect post-acceptance improvements, or any
corrections. The Version of Record is available online at: {\url{http://dx.doi.org/10.1007/s10714-024-03224-w}}.
 \end{acknowledgements}

 \bibliographystyle{spphys}       
\bibliography{adS4.bib} 

\end{document}